\documentclass[aps,pra,twocolumn,superscriptaddress]{revtex4-1}
\usepackage{amsmath}
\usepackage{amssymb}
\usepackage{graphicx}
\usepackage[export]{adjustbox}
\usepackage{color}
\usepackage{bm}
\usepackage{bigints}
\usepackage[colorlinks]{hyperref}
	\hypersetup{citecolor={red},urlcolor={blue}}
\usepackage{blkarray}
\usepackage{comment}
\usepackage{array}

\newcommand{\be}{\begin{eqnarray}}
\newcommand{\ee}{\end{eqnarray}}

\newcommand{\PRA}{\emph{Phys. Rev. A}}

\newcommand{\beq}{\begin{equation}}
\newcommand{\eeq}{\end{equation}}
\newcommand{\beqa}{\begin{eqnarray}}
\newcommand{\eeqa}{\end{eqnarray}}
\newcommand{\nn}{\nonumber}

\newcommand{\degr}{^{\circ}}

\newcommand{\ctamop}{\affiliation{Centre for Theoretical Atomic, Molecular and Optical Physics, School of Mathematics and Physics, Queen's University Belfast,
\\
University Road, Belfast, BT7 1NN, Northern Ireland, UK}}

\newcommand{\oumk}{\affiliation{Department of Physical Sciences, The Open University, Walton Hall, MK7 6AA Milton Keynes, UK}}

\newcommand\hlight[1]{}

\newcommand\rhlight[1]{}

\newcommand\ghlight[1]{}
\newcolumntype{C}[1]{>{\centering\arraybackslash}p{#1}}

\begin{document}

\title{Modeling tomographic measurements of photoelectron vortices in counter-rotating circularly polarized laser pulses}
\author{G. S. J. Armstrong}
\ctamop
\email[]{gregory.armstrong@qub.ac.uk}
\author{D. D. A. Clarke}
\ctamop
\author{J. Benda} 
\oumk
\author{J. Wragg}
\author{A. C. Brown}
\ctamop
\author{H. W. van der Hart}
\ctamop

\date{\today}
\begin{abstract}
Recent experiments [D. Pengel, S. Kerbstadt, L. Englert, T. Bayer, and M. Wollenhaupt, \href{https://journals.aps.org/pra/abstract/10.1103/PhysRevA.96.043426}{{\PRA} {\bf 96} 043426 (2017)}] have measured the photoelectron momentum distribution for three-photon ionization of potassium by counter-rotating circularly polarized 790-nm laser pulses. The distribution displays spiral vortices, arising from the interference of ionizing wavepackets with different magnetic quantum numbers. The high level of multidimensional detail observed in the distribution makes this an ideal case in which to demonstrate the accuracy of emerging theoretical techniques applicable to such problems. We use the \(R\)-matrix with time dependence approach to investigate this process. We calculate the full-dimensional photoelectron momentum distribution, and compare against a set of planar projections of this distribution previously measured in experiment. 
\end{abstract}


\maketitle

\section{Introduction}

Laser polarization provides a natural means of steering electron dynamics in atoms and molecules. The field of `programmable pulse shaping', in which the laser polarization can be tuned, provides avenues for multidimensional quantum control of ultrafast laser-driven electron dynamics \cite{silberberg2009,misawa2016,wollenhaupt2016}. Foundational techniques for laser polarization shaping \cite{brixner2001,brixner2004,suzuki2004,dudovich2004} exemplified control of the two-photon ionization yield in Rb, and multiphoton ionization yields in I$_2$ and K$_2$ molecules. Many similar demonstrations have been presented since these initial studies, and more will inevitably follow as  polarization-tailored pulses become more widespread \cite{kerbstadt2017a,kerbstadt2017b}. 

One particularly useful arrangement of pulses, namely a pair of time-delayed counter-rotating circular pulses, has become a prominent tool for influencing ionization dynamics in a range of atomic and molecular targets. Such a setup creates interference between wavepackets of differing magnetic quantum number, which is manifest in photoelectron momentum distributions in the form of spiral vortices \cite{pengel2017,pengel2017a,kerbstadt2019}. When the pulses are of different colors, this scheme also has important implications for high-order harmonic generation (HHG), and has long been recognized as a source of elliptically polarized attosecond pulse trains \cite{zuo1995,long1995,milosevic2000,fleischer2014,milosevic2015,kfir2015,lukas2015,dorney2017}.

In recent years, experimental analysis of such problems has been enhanced by the development of tomographic reconstruction techniques, with which the full three-dimensional momentum distribution of the ejected electrons may be built from a large set of planar projections \cite{wollenhaupt2009}. This technique, originally applied to atomic targets \cite{smeenk2009}, has also been extended to investigate arbitrarily-oriented polyatomic molecules \cite{hockett2010, wollenhaupt2013}.

{\em Ab initio} treatment of the dynamics driven by circularly polarized fields generally requires large-scale computing resources. This is primarily due to the necessary inclusion of magnetic sublevels, which quickly scales calculation size. For this reason, few {\em ab initio} methods have been developed with such pulses in mind. However, recent theoretical work has begun to uncover the dynamics of one- and two-electron systems in circularly polarized pulses. In particular, these studies have investigated ionization of hydrogen \cite{askeland2011,bauer2014,wu2016,ivanov2017}, single- and double-ionization of helium \cite{djiokap2014,djiokap2016,djiokap2017,donsa2019}, and double ionization of H\(_2\) \cite{djiokap2018,pindzola2017}. Studies of strong-field single-ionization of the H\(_2^+\), H\(_3^{2+}\), H\(_6^{5+}\), H\(_7^{6+}\), and HeH\(^{2+}\) molecular ions in bichromatic, circularly polarized pulses are also becoming available \cite{yuan2016,yuan2017,yuan2018}. These calculations have provided salient insights into the nature of electronic dynamics induced by arbitrarily-polarized fields. The response of larger molecules, such as benzene, to circular fields has also been investigated in the context of HHG using Floquet-based approaches \cite{averbukh2001} and time-dependent density-functional theory \cite{wardlow2016}.

However, when attempting to treat the interaction of multielectron systems with such fields, many theoretical methods rely on the single-active-electron (SAE) approximation \cite{kjeldsen2007,abu2011,ivanov2013,ivanov2014,mancusco2015,ilchen2017}, despite the suggestion of several studies \cite{weber2000,drescher2002,uiberacker2007,leone2014} that accurate capture of the dynamics requires a correlated, multielectron approach. The {\em ab initio} modeling of laser interactions with multielectron atoms is a demanding theoretical and computational task, requiring an accurate description of both the target electronic structure, and the ensuing strong-field ionization dynamics. Currently, one of the few methods capable of this is the \(R\)-matrix with time dependence (RMT) theory \cite{lampros2008,moore2011,clarke2018,brown2019}. RMT has been used to analyze the strong-field dynamics of a variety of atoms and ions, particularly in HHG by noble-gas atoms at near-IR wavelengths  \cite{hassouneh2014}, strong-field rescattering in F\(^-\) \cite{hassouneh2015}, as well as extreme-ultraviolet-initiated HHG in Ne \cite{brown2016} and Ar\(^+\)\cite{clarke2017}.  
Recently, RMT calculations of photoelectron momentum distributions for two-photon ionization of helium using counter-rotating circularly polarized pulses \cite{clarke2018} were compared against those from time-dependent close-coupling calculations \cite{djiokap2016}. A further study addressed the influence of the bound-electron magnetic quantum number on detachment yields from F\(^-\) in circularly polarized pulses \cite{armstrong2019}.

In this work, we use RMT to study resonance-enhanced multiphoton ionization (REMPI) of potassium by counter-rotating circularly polarized, 790-nm laser fields. At this wavelength, ionization occurs via a three-photon process, with a near-resonant \(4s\rightarrow 4p\) transition induced by the first photon absorption. We present a comparison of the measured photoelectron momentum distribution of Ref.\;\cite{pengel2017a} with that predicted using RMT. More specifically, we compare our calculated distributions with those measured in a set of planes in momentum space, which are ultimately used for tomographic reconstruction of the entire three-dimensional momentum distribution. The distributions vary strongly throughout momentum space, with spiral features dominating in the polarization plane and in several neighboring planes. In the planes perpendicular to the polarization plane, the distribution acquires a different character. Accurate capture of the measured features in full dimensionality represents a stringent and detailed test of the RMT method for atoms in arbitrarily-polarized laser fields, and enables a thorough assessment of its reliability in studies of photoelectron tomography.

\section{Theory}

\subsection{Numerical methods}


The RMT method for atoms and molecules in arbitrarily polarized laser fields is described in Refs.\;\cite{clarke2018} and \cite{brown2019}. RMT solves the multielectron time-dependent Schr\"{o}dinger equation in the electric dipole and non-relativistic approximations. The method divides configuration space into two distinct regions (inner and outer), according to radial distance of the outermost electron from the nucleus. The inner region is confined to small radial distances, such that all electrons are close to the target nucleus. In the outer region, a single outermost electron is allowed to extend from the end of the inner region to large radial distances --- typically a few thousand atomic units --- from the nucleus. In the outer region, the system can be considered as an electron subject to the long-range potential of the residual ion, as well as the laser field.

RMT uses a hybrid numerical scheme, that employs both basis-set and finite-difference techniques. In the inner region, the time-dependent, \((N+1)\)-electron wave function is expanded in \(R\)-matrix basis functions, with time-dependent expansion coefficients. The basis functions are generated from the wave functions of the residual-ion states, as well as a complete set of one-electron continuum functions describing the ionized electron. The outer-region wave function is constructed as a multichannel expansion, using residual-ion wave functions and radial wave functions of the ejected electron in each channel. We solve the resulting time-dependent Schr\"{o}dinger equation for the reduced radial wave function of the ejected electron, in each channel, using a finite-difference discretization scheme.

In both regions the laser-atom interaction is treated in the length gauge. This choice is merited by previous calculations \cite{hutchinson2010}, which demonstrated that the length gauge reduces the necessary level of atomic-structure detail by minimizing the role of short-range excitations near the nucleus.

\subsection{Electric field}
In this work, we compare against the measurements of Ref.\;\cite{pengel2017a} using 20-fs, 790-nm, \(5\times10^{10}\) W/cm\(^{2}\), counter-rotating, circularly polarized laser pulses. We adopt an electric field, polarized in the \(xy\) plane, of the form
\begin{align}
\bm{\mathcal E}(t)= & \; 
\frac{{\cal E}_0}{\sqrt{2}} \sin^2\left(\frac{\omega t}{2N_c}\right) 
\left[
\cos\omega t \;\hat{\bf x} - \sin\omega t \;\hat{\bf y}
\right]
\nn \\
& +
\frac{{\cal E}_0}{\sqrt{2}} \sin^2\left(\frac{\omega t'}{2N_c}\right) 
\left[
\cos\omega t' \;\hat{\bf x} + \sin\omega t' \;\hat{\bf y}
\right]
,
\label{efield}
\end{align}
where \({\cal E}_0\) is the peak electric field strength, \(N_c\) is the total number of cycles in each pulse, \(\omega = 1.56\) eV\;is the laser frequency,  and \(t'=t+\tau\), where \(\tau\) is the time delay between the maxima of the pulse envelopes. The peak laser intensity \(I_0\) is related to \({\cal E}_0\) using \(I_0 = c{\cal E}_0^2/4\pi\), where \(c\) is the speed of light in vacuum. The pulses are 20-fs in half-width, and last for \(N_c = 20\) laser cycles, consisting of a ten-cycle ramp-on and ramp-off. The majority of the calculations used a peak intensity of {\(5\times10^{10}\) W/cm\(^2\)}, in line with the measurements of Refs.\;\cite{pengel2017,pengel2017a}, though a number of calculations were performed at somewhat lower intensities, to investigate the effect of focal-volume averaging.

\section{Calculation parameters}

Within the inner region, the K\(^+\) ion is described using a set of Hartree-Fock \(1s\), \(2s\), \(2p\), \(3s,\) and \(3p\) orbitals for the K\(^+\) \(1s^2 2s^2 2p^6 3s^2 3p^6\) \(^1S^e\) ground state, employing the data of Clementi and Roetti \cite{clemroe}. We also include \(4s, 5s, 4p, 3d\) and \(4d\) orbitals, as well as a \(\bar{d}\) pseudo-orbital, originally obtained in Ref.\;\cite{berrington2006} using the CIV3 atomic structure code \cite{hibbert1975}. Using this set of orbitals, we calculate the ground state of K\(^+\) through a configuration-interaction calculation that includes the [Ne]\(3s^2 3p^6\) ground configuration, as well as excited configurations of the form \(3s^2 3p^5 nl\), \(3s^2 3p^4 (nl)^2\), and \(3s 3p^6 nl\).

\begin{table}[t]
    \centering
    \begin{tabular}{lC{1.5cm}C{1.5cm}}
    \hline\hline
            &  \multicolumn{2}{c}{Excitation energy (eV)} \\
      K state & Present & Ref.\;\cite{nist} \\
    \hline
     \(3p^6 4s\ ^2 S^e\)    & 0 & 0 \\
     \(3p^6 4p\ ^2 P^o\)    & 1.643 & 1.610 \\
     \(3p^6 5s\ ^2 S^e\)    & 2.647 & 2.607 \\
     \(3p^6 3d\ ^2 D^e\)    & 2.698 & 2.670 \\
     \(3p^6 5p\ ^2 P^o\)    & 3.078 & 3.063 \\
     \(3p^6 4d\ ^2 D^e\)    & 3.410 & 3.397 \\
     \(3p^6 6s\ ^2 S^e\)    & 3.419 & 3.403 \\
     \(3p^6 4f\ ^2 F^o\)    & 3.489 & 3.487 \\
     \(3p^6 6p\ ^2 P^o\)    & 3.598 & 3.596 \\
     \(3p^6 5d\ ^2 D^e\)    & 3.732 & 3.742 \\
     \(3p^6 7s\ ^2 S^e\)    & 3.734 & 3.754 \\
     \(3p^6 5f\ ^2 F^o\)    & 3.779 & 3.794 \\
     \hline\hline
    \end{tabular}
    \caption{Comparison between presently calculated and measured \cite{nist} excitation energies for selected field-free states of neutral K.}
    \label{entab}
\end{table}

The neutral K basis is then obtained by combining these target states with a set of continuum functions. These functions are generated using a set of 100 \(B\)-splines of order 9 for each available orbital angular momentum of the outgoing electron.
The maximum total angular momentum retained in the calculation is \(L_{\rm max} = 19\), a value which ensures a high level of convergence in the photoelectron momentum distributions for the three-photon ionization processes of interest in this work. We have performed a series of convergence checks using a larger basis with \(L_{\rm max}=29\), and observed negligible change in the resulting momentum distributions. The radial extent of the inner-region is set at 50 a.u., which suffices to confine the neutral K orbitals. We have verified that our results are well converged with respect to both the inner-region boundary radius, as well as the number of splines used in this range.

We obtain an ionization potential of 4.26 eV for the initial \(3s^2 3p^6 4s\) \(^2S^e\) K ground state, which is shifted to 4.34 eV to agree with the experimental value \cite{nist}. Excitation energies of higher-lying states are given in Table \ref{entab}. Additional excited states of the atom are not included, since they lie at least 20 eV above the ground state, and would require at least 14 790-nm (1.55 eV) photons for excitation. At the laser intensities used in this work, such high-order processes are of negligible probability.

In the outer region, the radial motion of the ejected electron is treated using a one-dimensional finite-difference grid, whose points are uniformly spaced by a distance of 0.08 a.u.. We adopt a fifth-order finite-difference scheme, which ensures a high degree of accuracy in describing the spatial properties of the ionized-electron wavepacket. The outer-region grid extends to large radial distances, in this case up to 4800 a.u., enabling the asymptotic characteristics of the ionized-electron wavefunction to be faithfully determined. The photoelectron momentum distribution is, of course, particularly sensitive to these characteristics.

To propagate the wave function in time, we use an Arnoldi propagator of order 8 \cite{moore2011}, with a timestep \(\delta t = 0.01\) a.u., to ensure a highly-accurate final wave function. To provide ionizing wavepackets with sufficient time to reach the outer region, the wave function is propagated for a significant length of time after the pulse has terminated. In all cases, we set a total propagation time of 7000 a.u. (close to 170 fs). This setting provides ample time for even low-energy electrons to reach the outer region.

Following the time propagation, we obtain the radial part of the ejected-electron wave function in each channel, once its angular momentum has been decoupled from that of the residual system \cite{vdh2008,lysaght2009}. The wave function is then transformed, for \(r > 50\) a.u., into the momentum representation by performing a Fourier transform. Analysis of the channel wave functions shows that even the lowest-energy wavepackets have reached radial distances of at least \(r > 50\) a.u.\;by the final propagation time, and possess continuum character, indicating that the full photoelectron wave function is faithfully described in momentum space. In particular, by the end of the time propagation, the dominant \(f_{\pm3}\) photoelectron wavepackets are localized between radial distances of 300 a.u. and 2000 a.u..  Further details on this procedure are given in Appendix \ref{momdisapp}.



\begin{figure}
    \centering
    \includegraphics[width=\columnwidth,trim={0 0cm 0 0cm}]{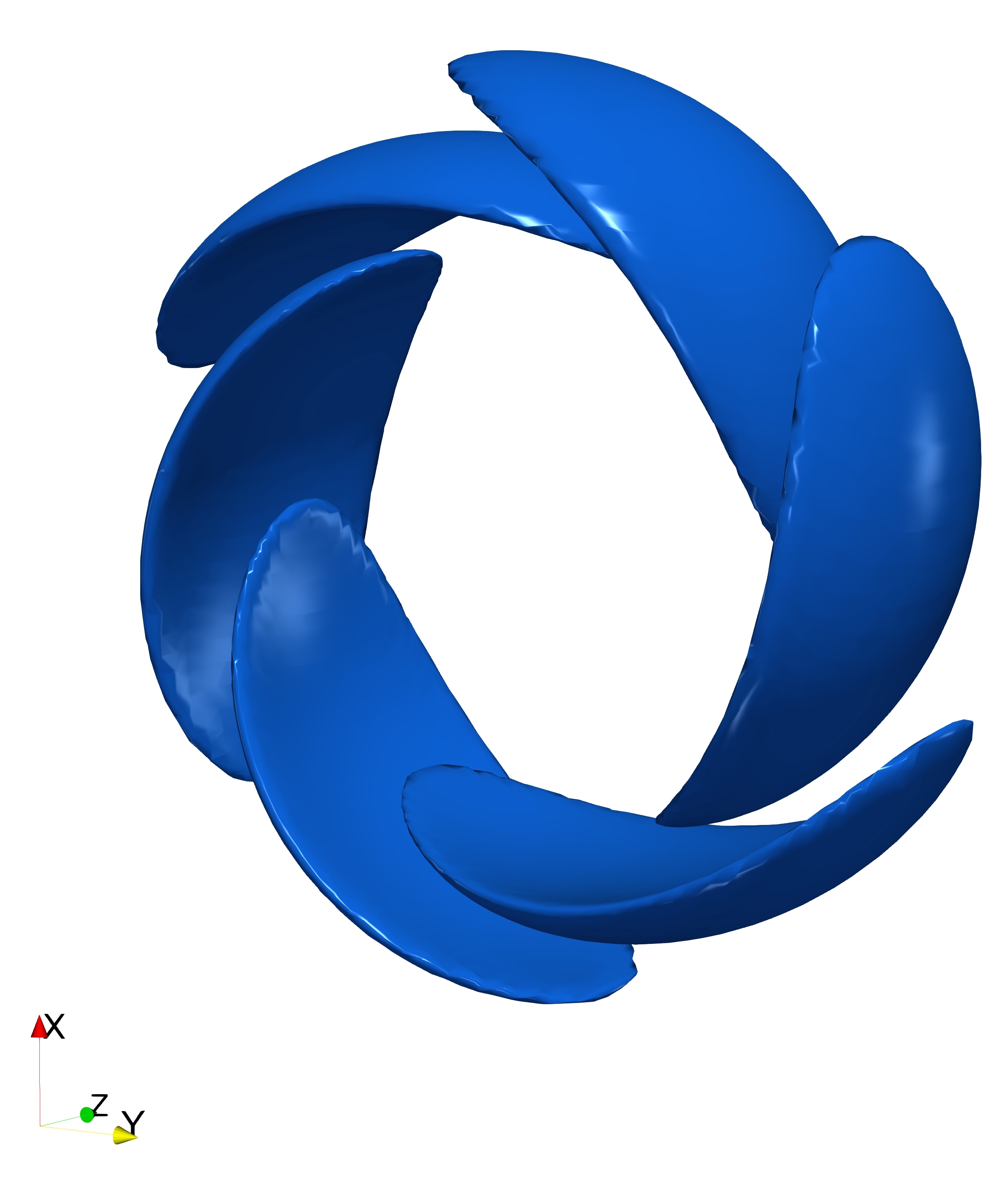}
    \caption{Full-dimensional energy-space isosurface calculated by RMT for a pair of 20-fs, 790-nm, \(5\times10^{10}\) W/cm\(^2\) pulses with a relative delay of 40 fs.}
    \label{fig:3dplot}
\end{figure}


\begin{figure}[t]
$
\begin{array}{cc}
(a)\ \text{measurement,}\ xy \ \text{plane} & (b)\ \text{RMT,}\ xy \ \text{plane},\ \tau = 40\ \text{fs}  
\\
{\centering\includegraphics[width=0.5\columnwidth,valign=c,trim={1cm 2cm 1cm 1cm}]
{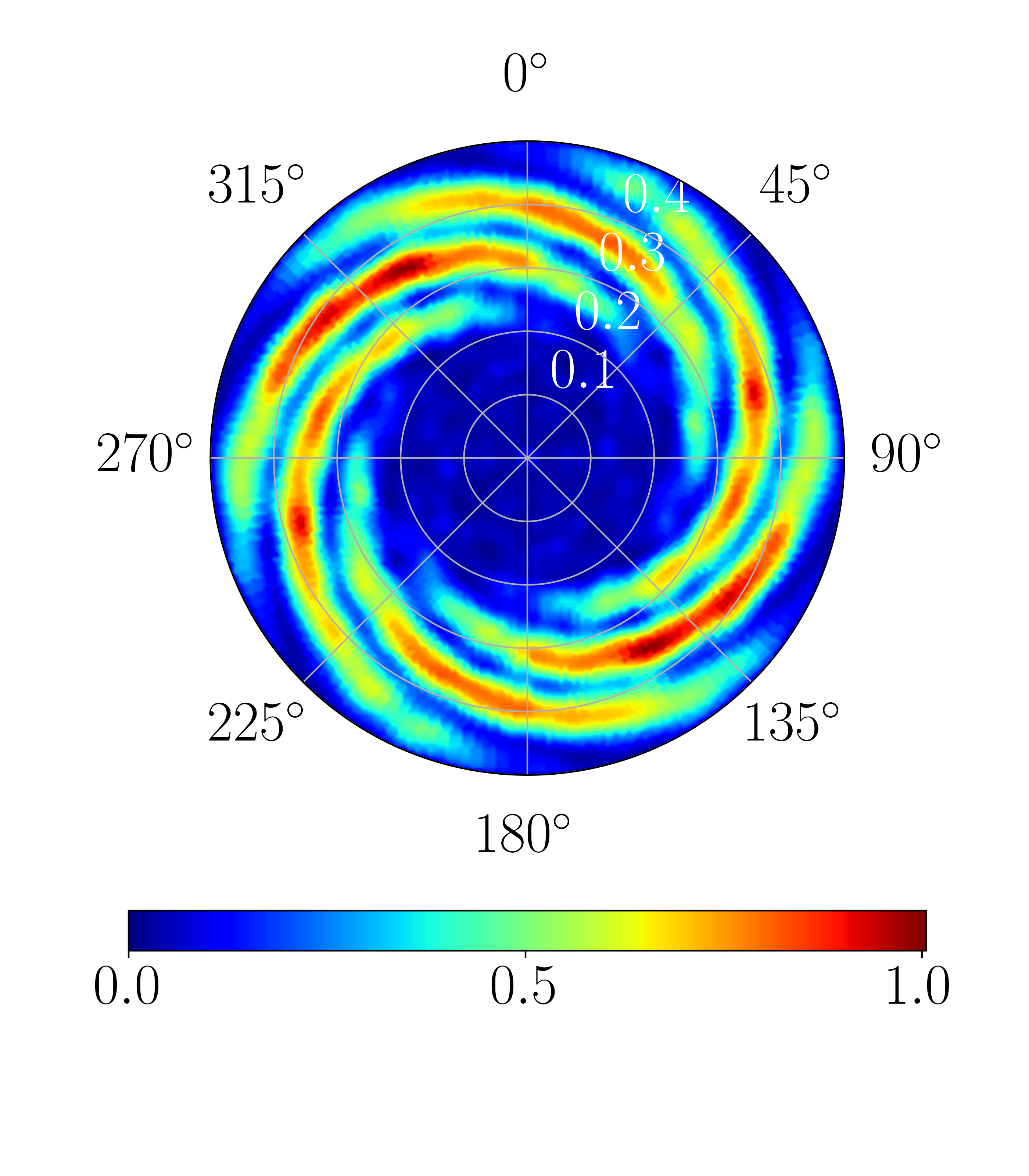}
}
&
{\centering\includegraphics[width=0.5\columnwidth,valign=c,trim={1cm 2cm 1cm 1cm}]
{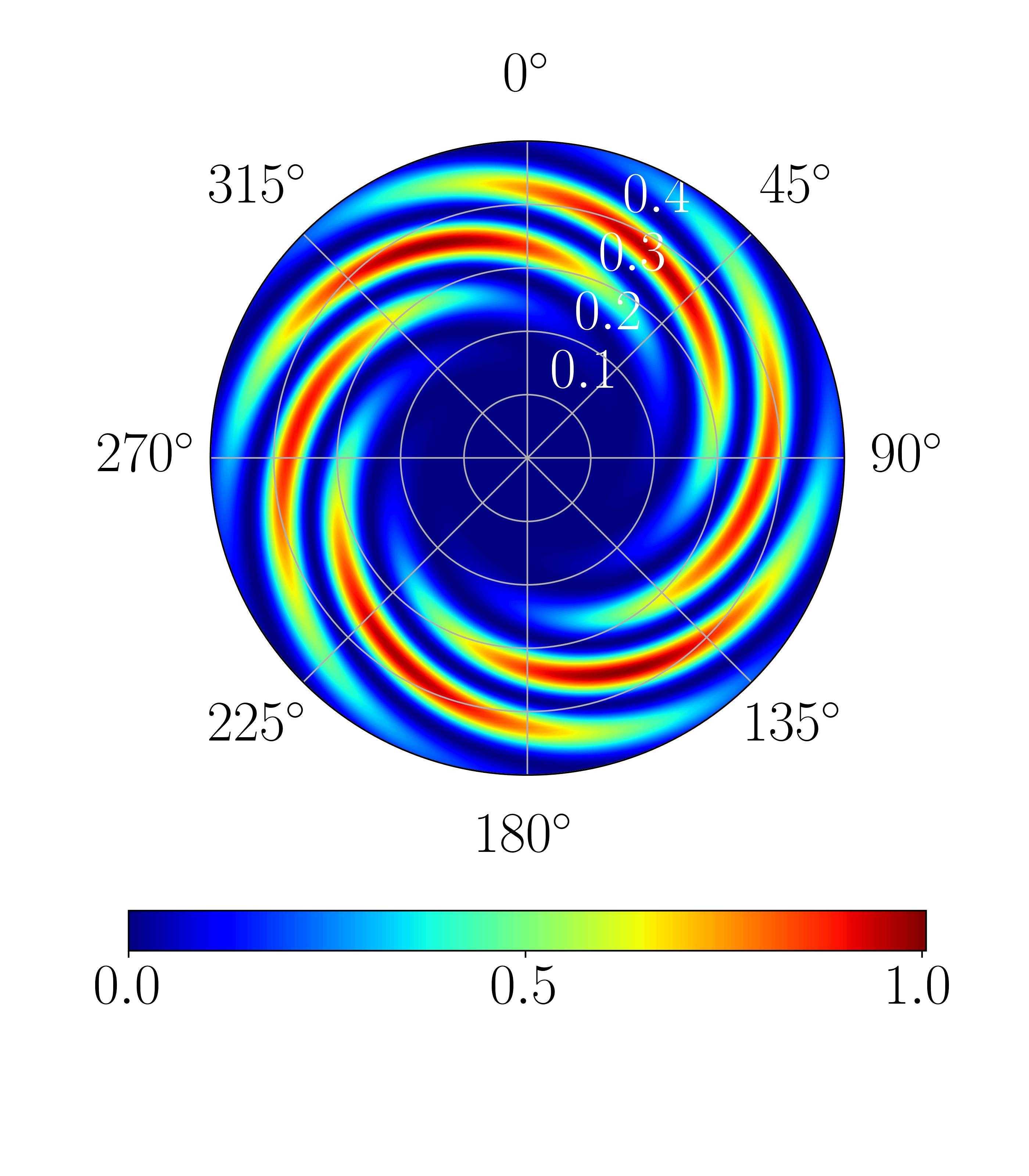}
}
\\
(c)\ \text{RMT,}\ xy \ \text{plane},\ \tau = 38\ \text{fs} & (d) \ \text{RMT,}\ xy \ \text{plane}, \ \tau = 42\ \text{fs}  
\\
{\centering\includegraphics[width=0.5\columnwidth,valign=c,trim={1cm 2cm 1cm 1cm}]
{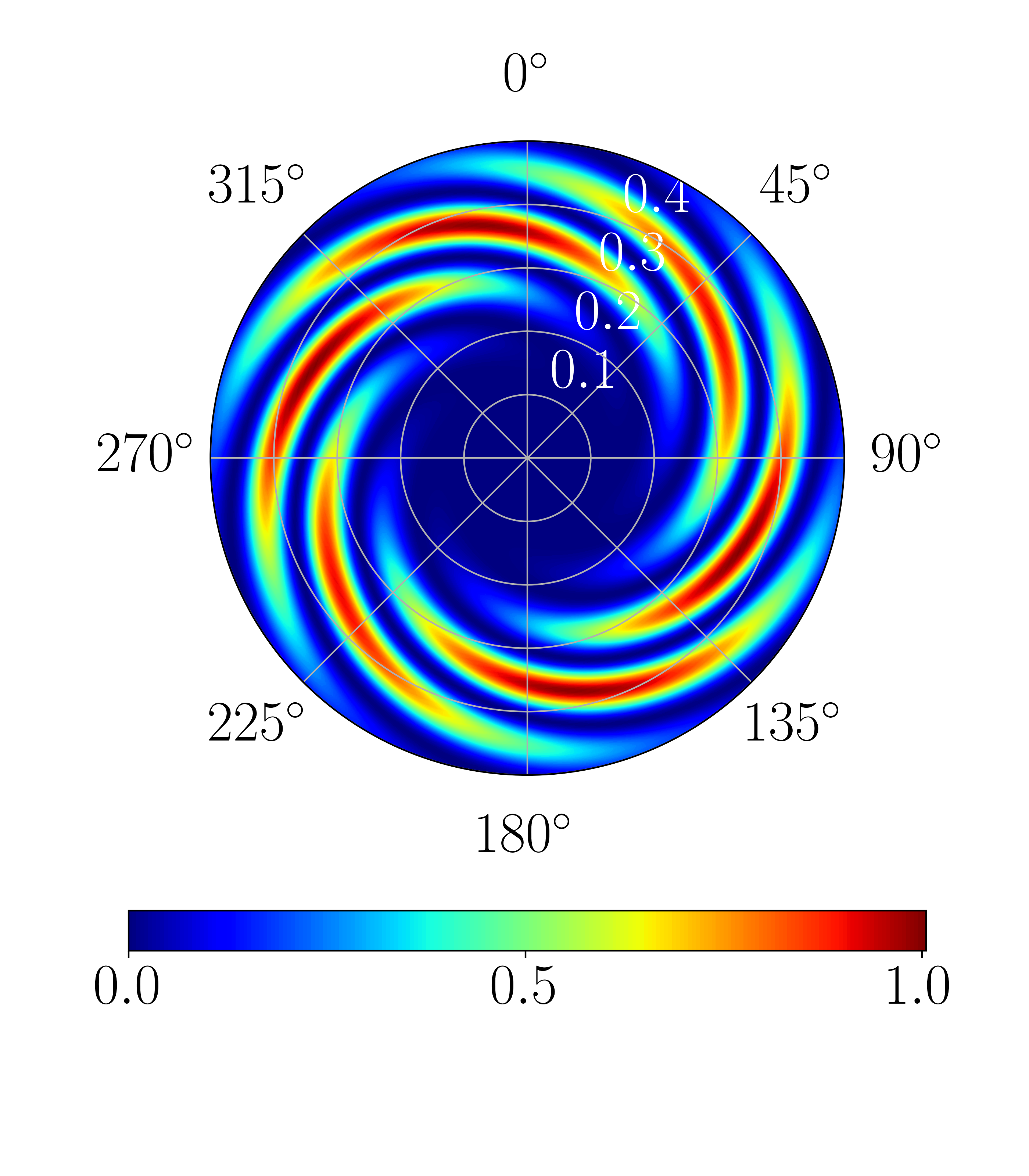}}
&
{\centering\includegraphics[width=0.5\columnwidth,valign=c,trim={1cm 2cm 1cm 1cm}]
{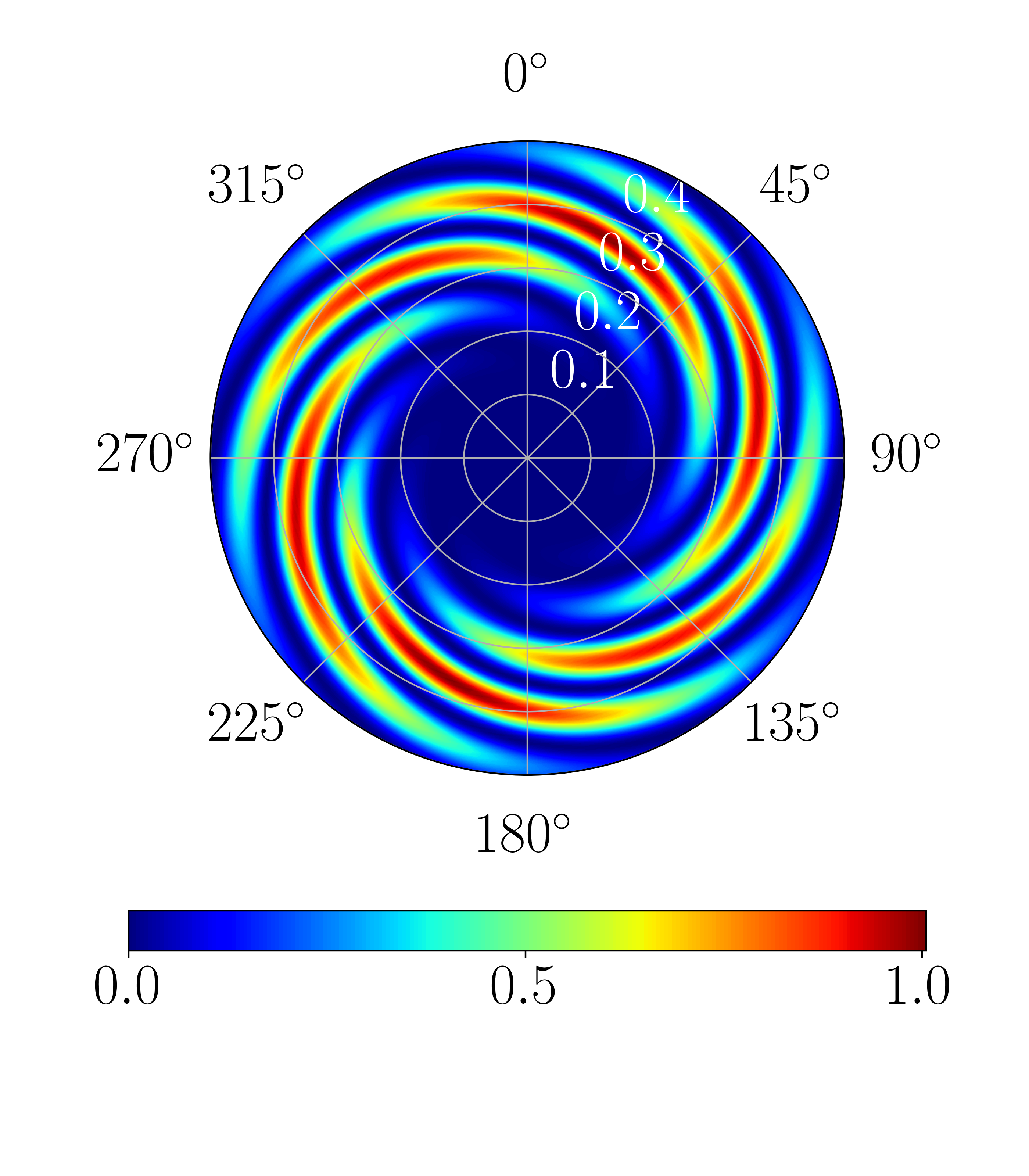}}
\end{array}$
\caption{Normalized photoelectron momentum distributions in the polarization plane, (a) as measured in Ref.\;\cite{pengel2017a}, and as calculated by RMT using time delays of (b) 40 fs, (c) 38 fs, and (d) 42 fs, following ionization of K, initiated by a pair of 20-fs, 790-nm, $5 \times 10^{10}$ W/cm$^{2}$, counter-rotating circularly polarized laser pulses. Note that, as in Ref.\;\cite{pengel2017a}, the azimuthal angle \(\phi\) is traversed in a clockwise sense in each figure, starting from \(0^{\circ}\) in the vertical direction. Values along the radial axes indicate photoelectron energy in eV.}
\label{exp20fscomp}
\end{figure}

\section{Results and Discussion}

The spiral distributions induced by counter-rotating circularly polarized pulses can now be observed experimentally \cite{pengel2017,pengel2017a,kerbstadt2019} in full dimensionality, through tomographic reconstruction techniques. In Ref.\;\cite{pengel2017a}, such a distribution was measured for ionization of K driven by a pair of 790-nm, 20-fs, $5\times10^{10}$ W/cm$^2$, counter-rotating circularly polarized laser pulses. Fig.\;\ref{fig:3dplot} presents the result of an RMT calculation that aims to replicate this experimental capability. The figure shows the energy-space isosurface for ionization of K, driven by a pair of counter-rotating laser pulses with a time delay of 40 fs between them. The distribution assumes the form of a six-arm, helical vortex structure, with a two-dimensional spiral cross section in the laser polarization plane, and a finite extension across neighboring planes. The source of these features is well known, namely the interference between outgoing electrons with different magnetic quantum numbers. As outlined in Ref.\;\cite{pengel2017a}, they arise in this case from interference between $f_3$ and $f_{-3}$ continuum electrons. An additional REMPI process is possible, where the \(4p_{-1}\) state is populated by one photon absorption from the first pulse, and the second pulse then induces two-photon ionization from this state, leading to the ejection of an \(f_{1}\) electron. However, at low intensity this pathway is strongly suppressed, with the yield of \(f_{\pm3}\) electrons exceeding that of \(f_1\) electrons by two orders of magnitude. Using a perturbative analysis, and considering only \(f_{\pm3}\) photoelectrons, the full-dimensional energy-space distribution \(P(E,\theta,\phi)\) takes the form (similar to Eq. (4) of Ref.\;\cite{pengel2017a})
\begin{equation}
    P(E,\theta,\phi) = 
    g(E)\sin^6\theta\left[1-\cos(6\phi+E\tau)\right],
    \label{pertex}
\end{equation}
where \(E\) is the photoelectron energy, \(\tau\) is the time delay between the two pulses, and \(g(E)\) is the energy distribution.

Figure \ref{exp20fscomp} presents slices of this distribution in the laser polarization plane. The measured distribution, shown in Fig.\;\ref{exp20fscomp}(a), was first presented in Ref.\;\cite{pengel2017a}, where it was also demonstrated, by a simple perturbative analysis, that this distribution is sensitive to the time delay between the laser pulses. Here, we investigate this sensitivity by performing a set of three calculations, scanning over time delays \(\tau\) between 38 fs and 42 fs. The resulting distributions are shown in Fig.\ref{exp20fscomp}(b)-(d).

The overall agreement between the calculated and measured distributions appears good. Each arm is confined to an annular energy range, from approximately 0.2 eV to around 0.5 eV. In the measured distribution, the low-energy tails of the spiral arms begin at around \(\phi \approx (40+60n)^{\circ}\), for \(n=0-5\), with each arm sweeping out around 150\(^{\circ}\). The RMT calculation of Fig.\ref{exp20fscomp}(b), in which \(\tau = 40\) fs, displays very similar features, with both the energy span and angular extent of the spirals closely matching those of the measurement. However, a small angular offset relative to the measurement is visible, with the arms of the calculated distribution appearing at approximately \(\phi\approx(45+60n)^{\circ}\), for \(n=0-5\).

Figs.\;\ref{exp20fscomp}(c) and (d) show analogous distributions, obtained for values of \(\tau\) within the experimental error estimate. Clearly, the time delay affects the orientation of the spirals, shifting it by around 45\(^{\circ}\) in an anti-clockwise direction as the delay is decreased from 40 fs to 38 fs, and causing a similar clockwise shift of 45\(^{\circ}\) as the delay is increased to 42 fs. The distribution shown in Fig.\ref{exp20fscomp}(d), in which \(\tau=42\) fs, displays an orientation that very closely resembles that of the measurement. 

The sensitivity to time delay observed here is an important factor. Measured time delays may have uncertainties of around 5 -- 10\%. A natural method of reducing such uncertainties, and thereby of calibrating the measured delay, is through examination of the distributions calculated at a number of individual time delays. The variations in the calculated distributions of Fig.\;\ref{exp20fscomp} suggest that the measured time delay may be further fine-tuned within a range centered close to 42 fs. 

In the calculated distributions, the spirals present a strong six-fold symmetry, with nearly commensurate peaks. This symmetry arises partly due to the large temporal separation of the two pulses, which negates any significant ejection of \(f_{\pm1}\) electrons due to the overlap of the two pulses. Furthermore, at the intensity used here, the near-resonant \(4s \rightarrow 4p\) transition is not strongly driven, meaning that ejection of an \(f_{1}\) electron is suppressed. However, in all cases, two of the spiral arms attain a marginally lower magnitude than the others. For example, for \(\tau = 40\) fs, the smaller maxima appear at around 100\(^{\circ}\) and 280\(^{\circ}\). Such deviations from six-fold symmetry are due to weak emission of an \(f_1\) electron via the REMPI pathway. 

The angular locations of the maxima in the RMT calculations closely resemble those appearing in experiment. However, a minor degree of asymmetry is visible in the measured distribution, with the two peaks around $\phi\approx 135^{\circ}$ and $315^{\circ}$ rising noticeably higher than the others. We note, however, that aside from these rather localized maxima, the measured distribution presents a strong symmetry, particularly in the spiral tails, that is very reminiscent of that displayed in the calculations.

Clearly, the symmetry properties of the distribution are largely unaffected by time delays in the range shown, with a high degree of six-fold symmetry persistent in all cases. Further calculations (not shown) demonstrate that an asymmetry is not apparent in the distribution until the time delay is reduced to around 30 fs, which is well outside the error estimate of the measured delay. This asymmetry induced by ionization pathways involving both pulses is well known, and has been observed and analyzed in previous work \cite{djiokap2016,clarke2018}.

\begin{figure}[t]
$
\begin{array}{cc}
(a)\ I = 5\times10^{10}\ \text{Wcm}^{-2} & (b)\ \text{intensity-averaged}
\\
{\centering\includegraphics[width=0.5\columnwidth,trim={1cm 2cm 1cm 1cm},valign=c]
{polar-momdis-K-790nm-20fs-051010-L19-momentum-rskip0-xyplane-dbar-nofbar_d42fs.png}
}
&
{\centering\includegraphics[width=0.5\columnwidth,trim={1cm 2cm 1cm 1cm},valign=c]
{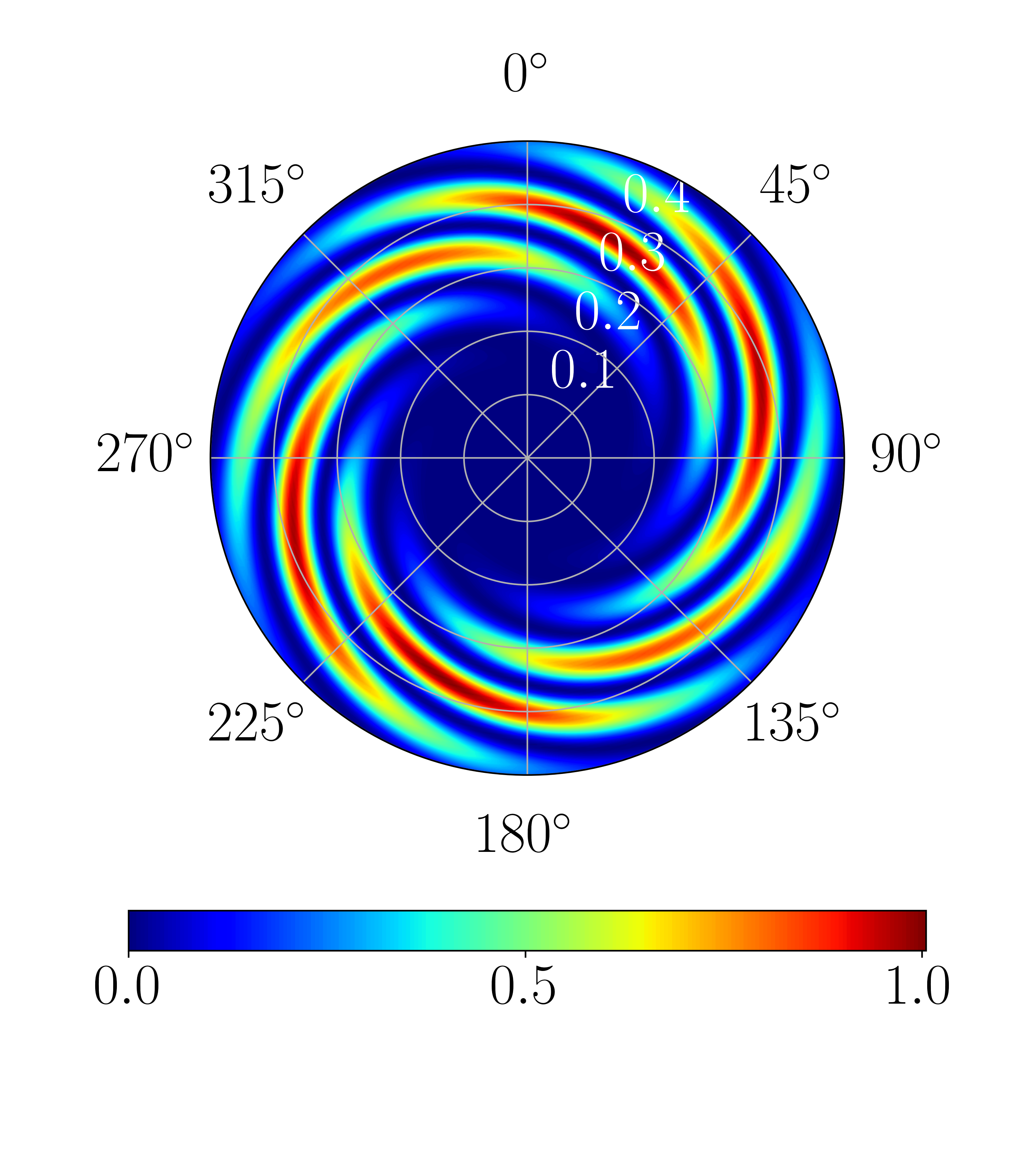}
}
\end{array}$
\caption{Normalized photoelectron momentum distributions in the \(xy\) plane 
at a peak laser intensity of \(5\times10^{10}\) W/cm\(^2\), compared to the distribution obtained after intensity averaging for a fixed time delay of 42 fs.}
\label{idep}
\end{figure}

\begin{figure}[ht]
$
\begin{array}{cc}
 (a)\ \text{measurement,}\ yz \ \text{plane}  & (b)\ \text{RMT,}\ yz \ \text{plane}, \tau = \text{40 fs}
\\
{\centering\includegraphics[width=0.5\columnwidth,trim={1cm 2cm 1cm 1cm},valign=c]{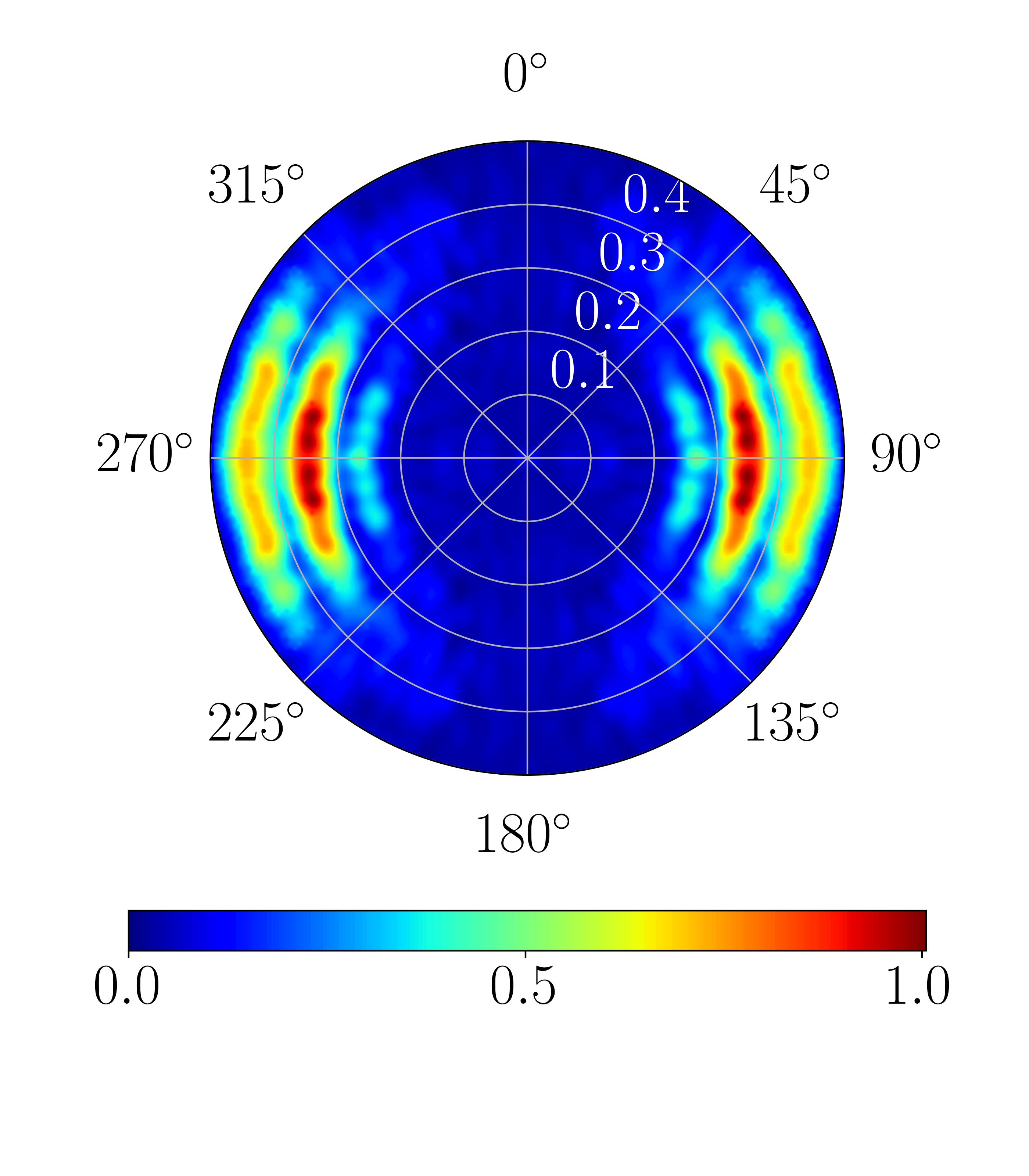}}
&
{\centering\includegraphics[width=0.5\columnwidth,trim={1cm 2cm 1cm 1cm},valign=c]
{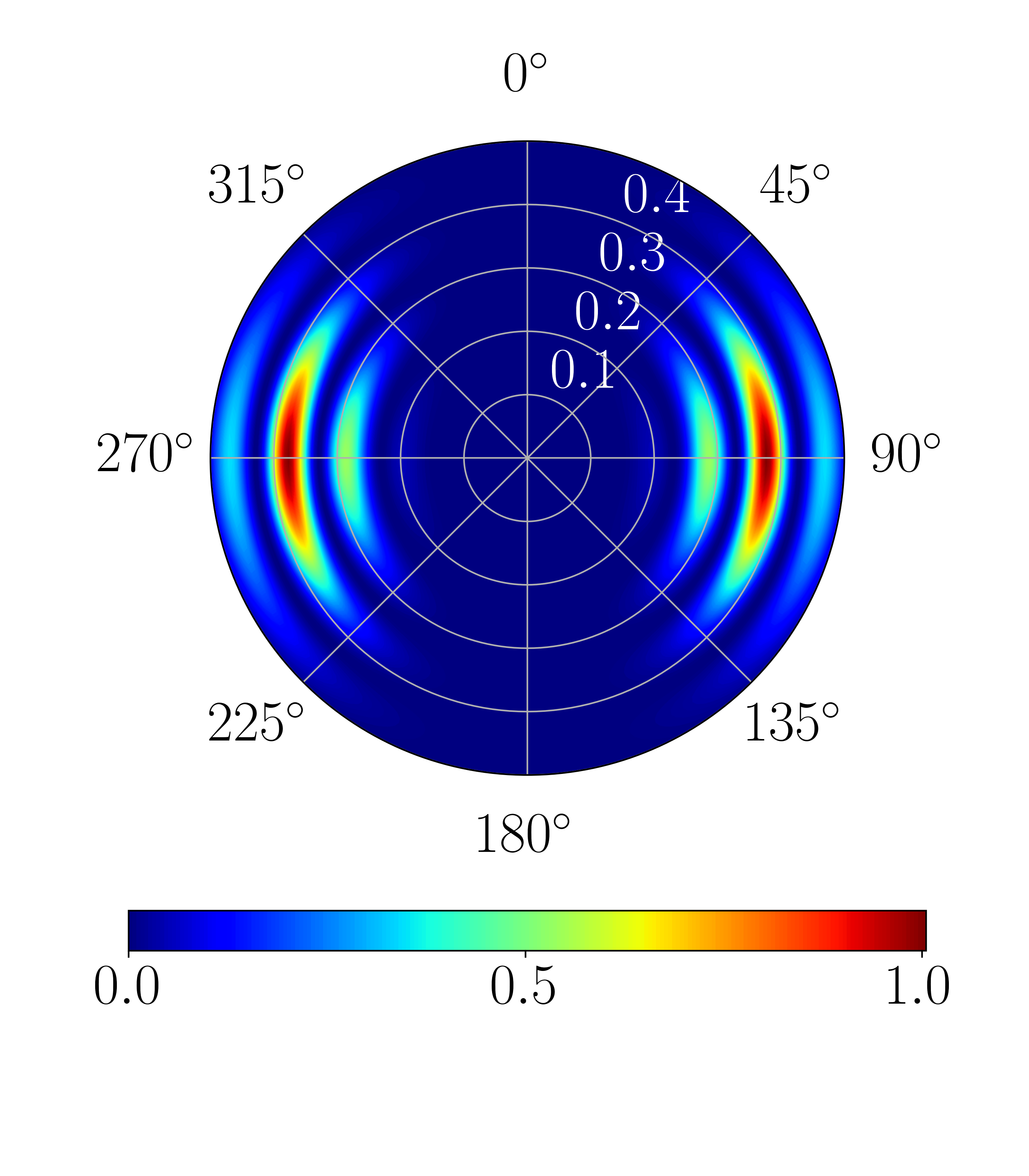}
}
\\
(c)\ \text{RMT,}\ yz \ \text{plane}, \tau = \text{38 fs} & (d)\ \text{RMT,}\ yz \ \text{plane}, \tau = \text{42 fs}
\\
{\centering\includegraphics[width=0.5\columnwidth,trim={1cm 2cm 1cm 1cm},valign=c]{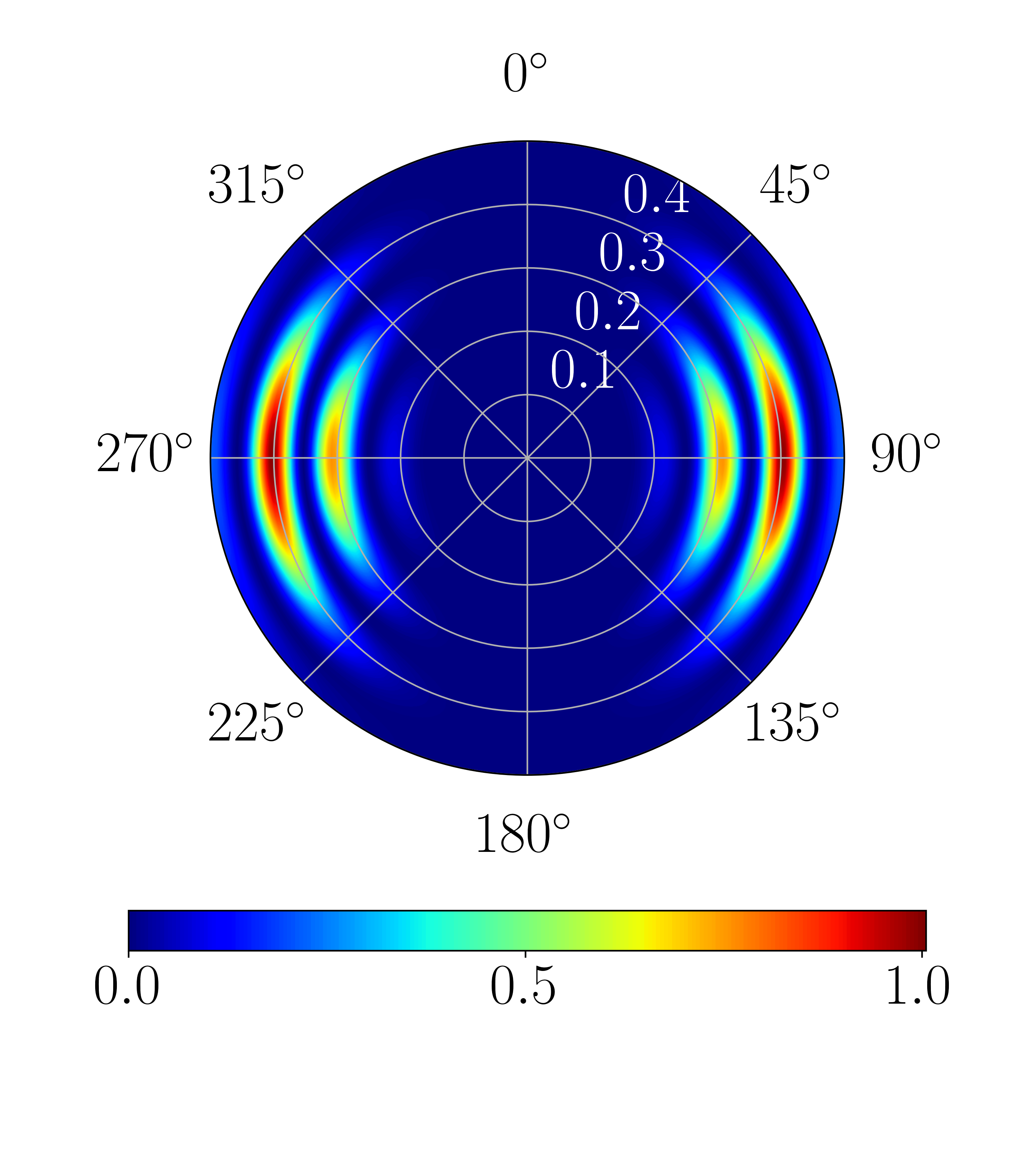}
}
&
{\centering\includegraphics[width=0.5\columnwidth,trim={1cm 2cm 1cm 1cm},valign=c]{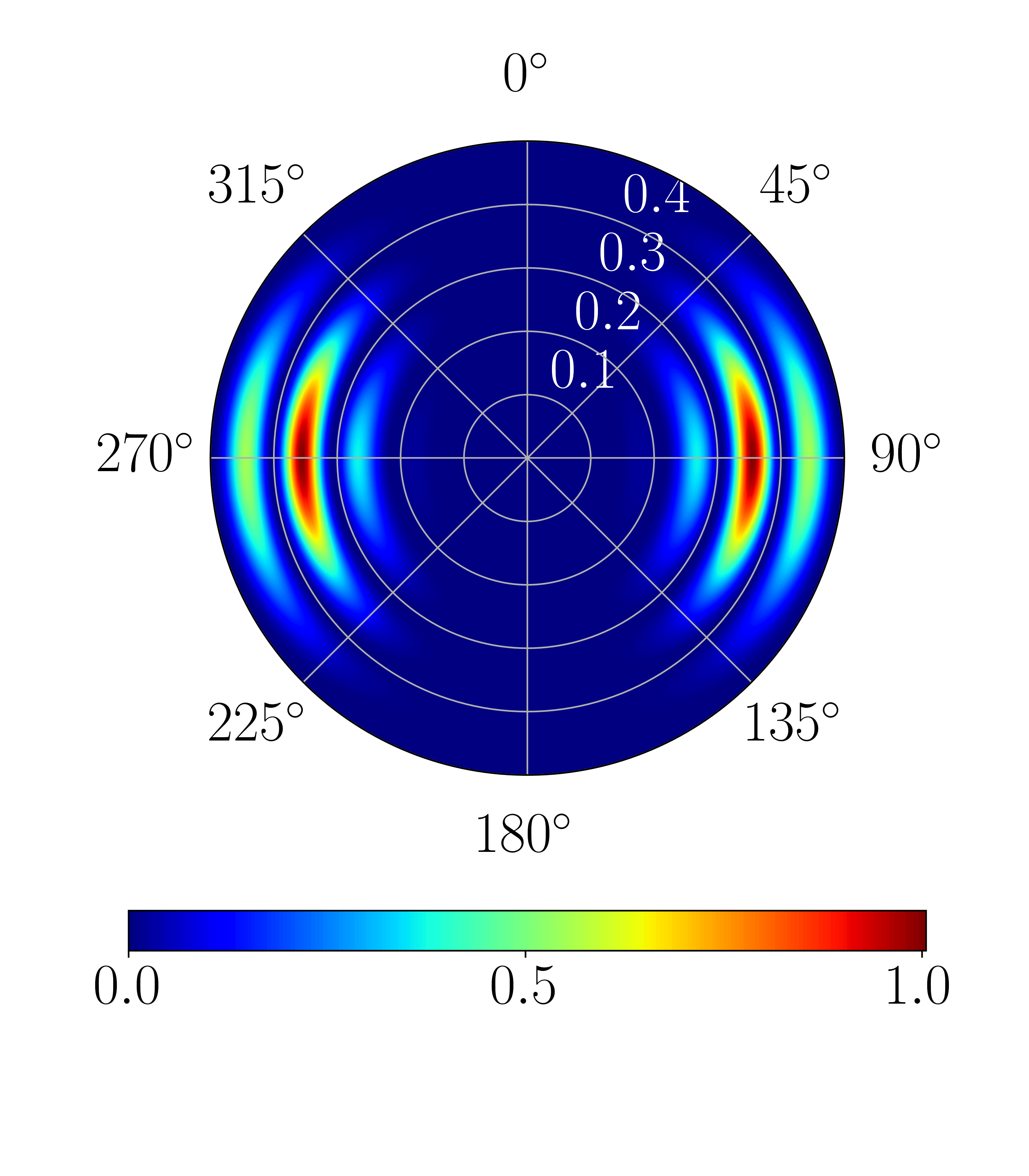}
}
\end{array}$
\caption{Normalized photoelectron momentum distributions in the \(yz\) plane, (a) as measured and (b)-(d) as calculated by RMT, following ionization of K, initiated by a pair of 20-fs, 790-nm, $5 \times 10^{10}$ W/cm$^{2}$, counter-rotating circularly polarized laser pulses, using time delays of (b) 40 fs, (c) 38 fs, and (d) 42 fs. In each figure the polar angle \(\theta\) is traversed in a clockwise direction, starting from 0\(\degr\) in the upward vertical direction.}
\label{exp20fsyz}
\end{figure}

\begin{figure}[ht]
$
\begin{array}{cc}
 (a)\ \text{measurement,}\ xz \ \text{plane}  & (b)\ \text{RMT,}\ xz \ \text{plane}, \tau = \text{40 fs}
\\
{\centering\includegraphics[width=0.5\columnwidth,trim={1cm 2cm 1cm 1cm},valign=c]{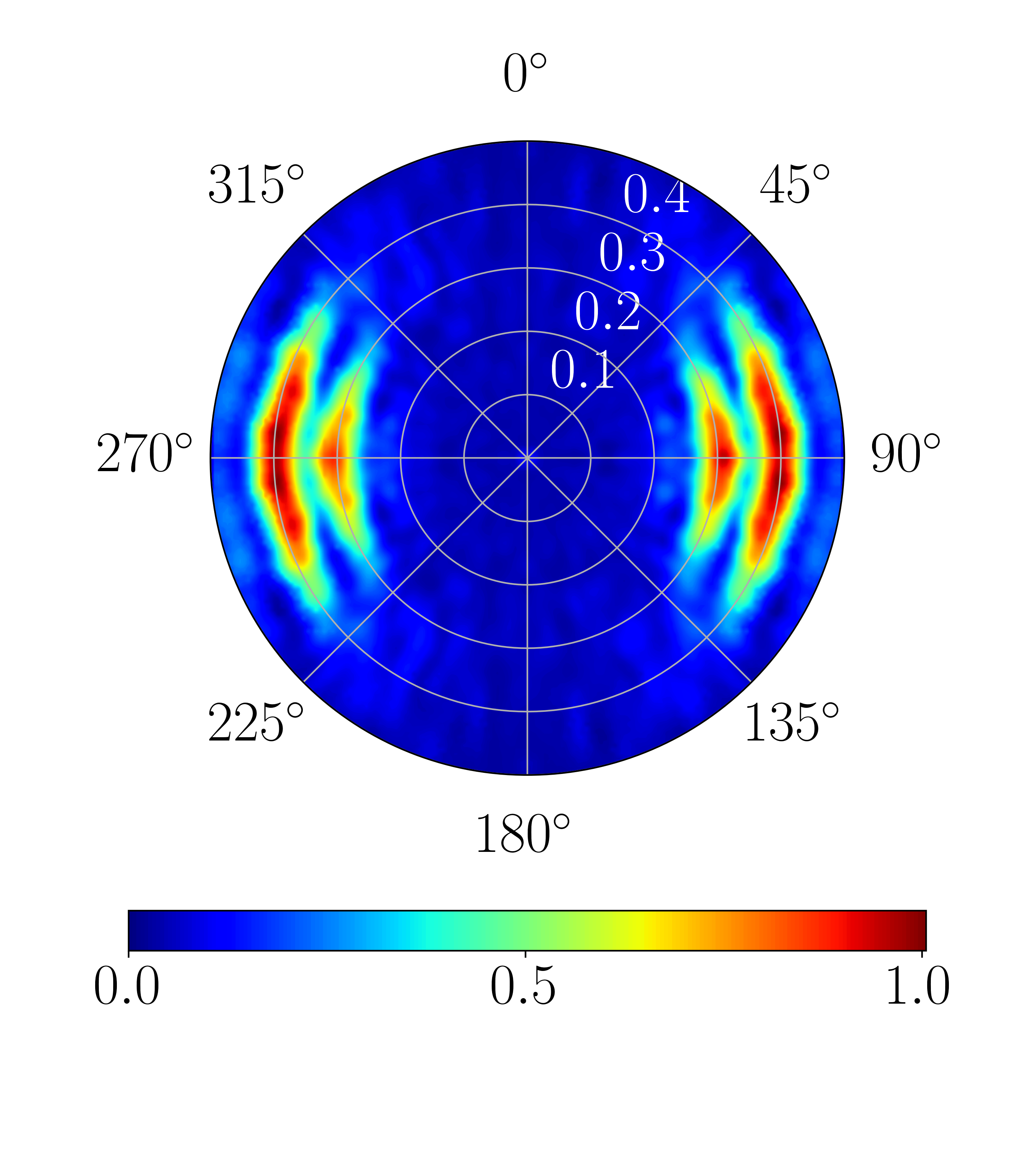}}
&
{\centering\includegraphics[width=0.5\columnwidth,trim={1cm 2cm 1cm 1cm},valign=c]
{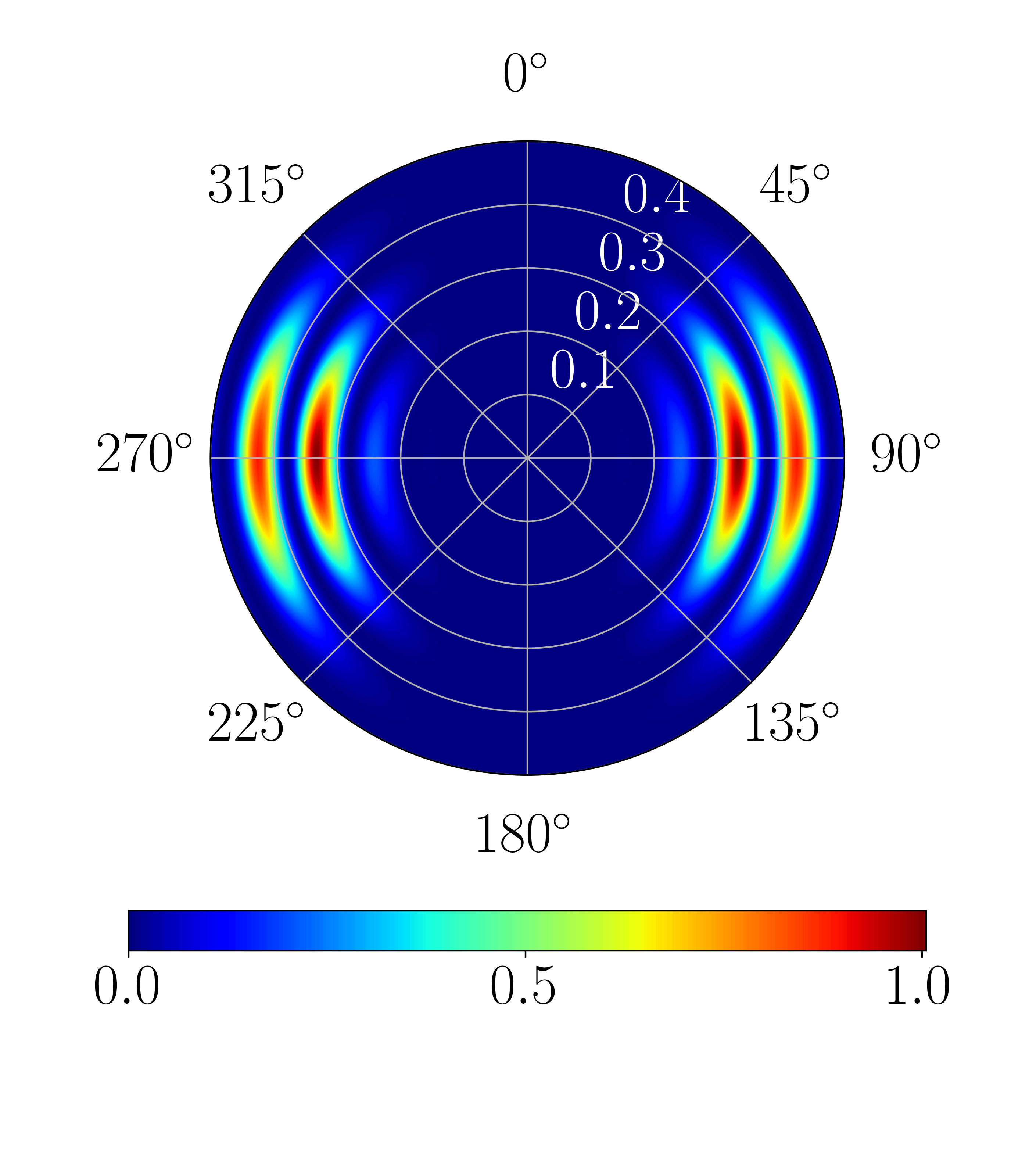}
}
\\
(c)\ \text{RMT,}\ xz \ \text{plane}, \tau = \text{38 fs} & (d)\ \text{RMT,}\ xz \ \text{plane}, \tau = \text{42 fs}
\\
{\centering\includegraphics[width=0.5\columnwidth,trim={1cm 2cm 1cm 1cm},valign=c]
{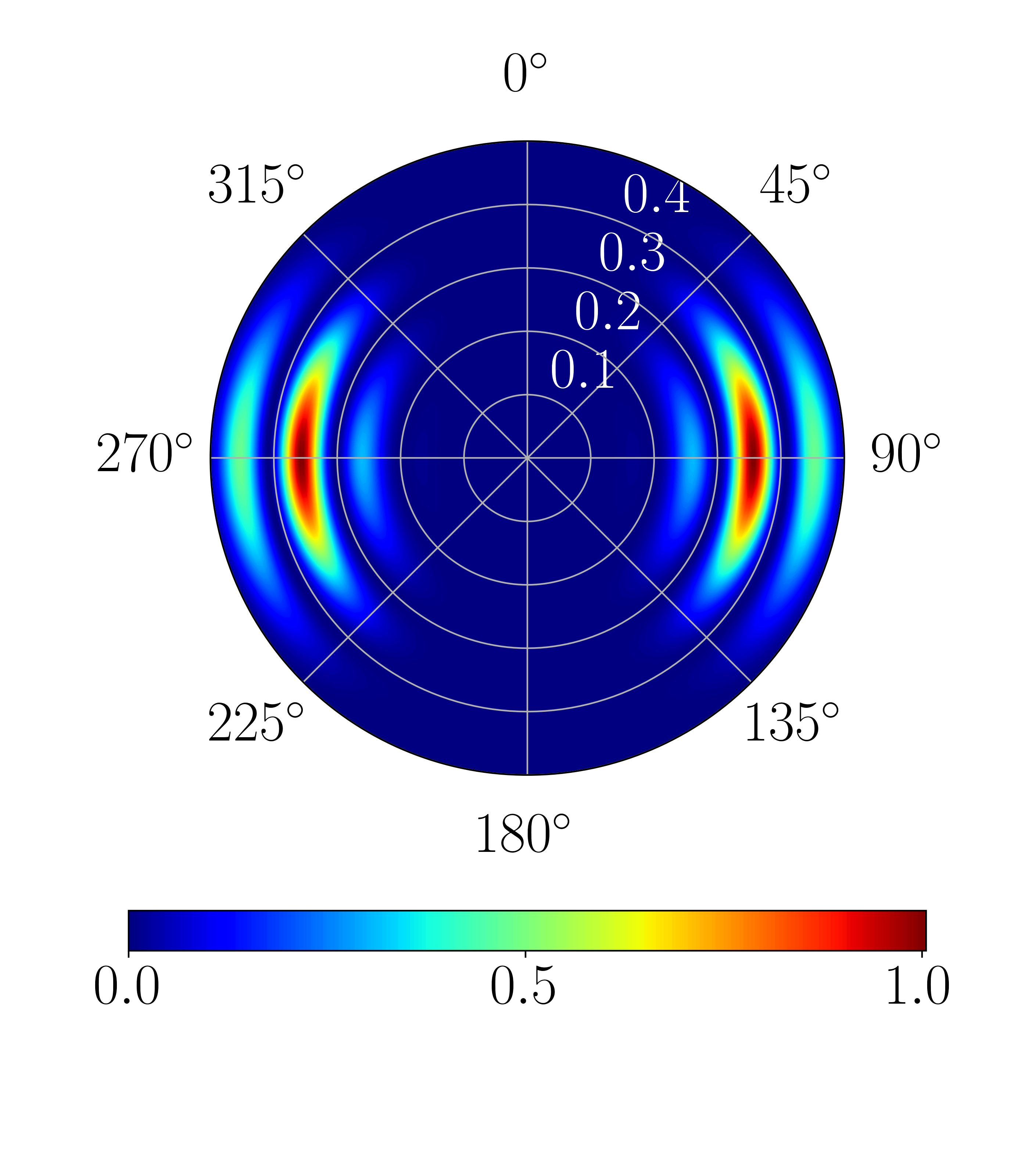}
}
&
{\centering\includegraphics[width=0.5\columnwidth,trim={1cm 2cm 1cm 1cm},valign=c]
{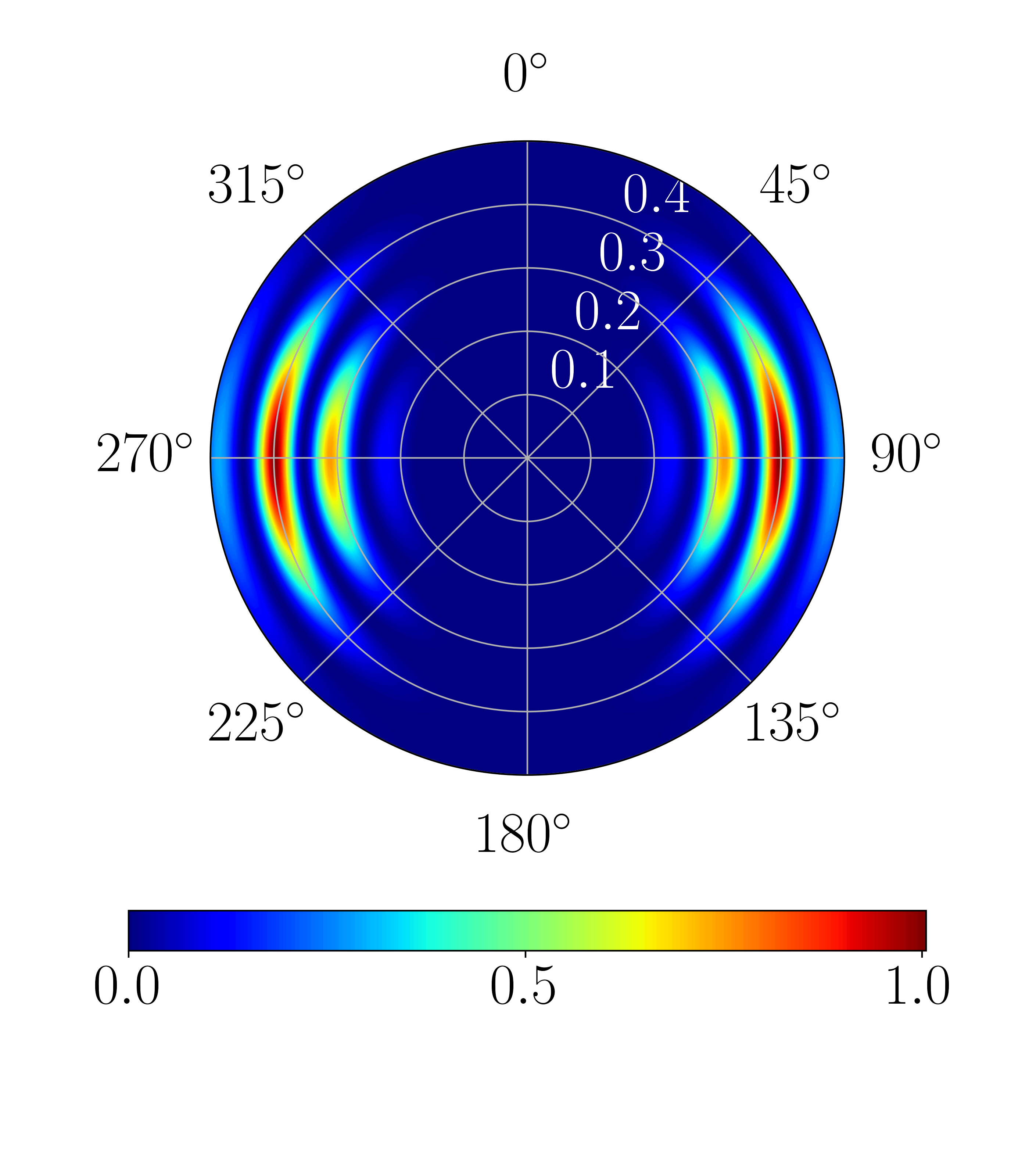}
}
\end{array}$
\caption{Normalized photoelectron momentum distributions in the \(xz\) plane, (a) as measured and (b)-(d) as calculated by RMT, following ionization of K, initiated by a pair of 20-fs, 790-nm, $5 \times 10^{10}$ W/cm$^{2}$, counter-rotating circularly polarized laser pulses, using time delays of (b) 40 fs, (c) 38 fs, and (d) 42 fs.}
\label{exp20fsxz}
\end{figure}

We have also investigated the pulse-length dependence of the distributions. We find that the pulse duration mainly affects the length of the spiral arms, and has a negligible effect on the positions of the maxima along each of the arms. As the laser pulse is shortened, the annulus to which the spirals are confined broadens in energy, and retracts as pulse length increases. The width of the bounding annulus is therefore directly linked to the bandwidth of the laser pulse, in a similar fashion to typical above-threshold ionization peaks in photoelectron spectra.

At this point we must consider whether or not a comparison between calculations at a single laser intensity and experiment is valid. In strong-field processes, observables such as photoelectron spectra may depend sensitively on the laser intensity, and therefore focal-volume averaging is necessary for a rigorous comparison of theoretical data with experiment. However, despite the near-infrared (790 nm) wavelength used here, the process of interest involves only three photon absorptions from each pulse, and the energy distribution consists of a single significant peak, in contrast to typical above-threshold ionization spectra containing multiple peaks spanning a wide range of energies. Furthermore, in this comparison we consider the normalized photoelectron distribution. While we expect the laser intensity to affect the total ionization yield, normalization of the distribution removes this obvious source of intensity dependence. It is therefore unclear if focal-volume averaging is warranted in this case. To proceed, we investigate the effect of intensity (focal-volume) averaging on our calculated distributions.

Fig. \ref{idep} compares the normalized distribution at a fixed peak laser intensity of {\mbox{\(5\times10^{10}\) W/cm\(^2\)}}, with that obtained following an averaging over a set of peak intensities, for a time delay of \(\tau = 42\) fs. Evidently, the distributions are weakly sensitive to the laser intensity averaging. 
The orientation of the spirals is largely unaffected, and their position and width in energy are also unaltered. In fact, the main intensity-dependence appears to be confined to the central part of the spiral arms that extend from around 120\(^{\circ}\) to 180\(^{\circ}\), and from 300\(^{\circ}\) to 0\(^{\circ}\). Here, the intensity-averaged yield is a few percent lower than that obtained at the peak intensity. An even weaker sensitivity is visible in the arm whose largest magnitudes extend from around 45\(\degr\) to 90\(\degr\).  Given the weakness of the intensity dependence, we proceed with the comparison at a single laser intensity.


We now compare slices of the photoelectron momentum distributions through planes away from the laser polarization plane. We begin with Fig.\;\ref{exp20fsyz},  where we compare measured (Fig.\;\ref{exp20fsyz}(a)) and calculated (Figs.\;\ref{exp20fsyz}(b)-(d)) distributions in the \(yz\) plane (perpendicular to the laser polarization plane). The distributions are plotted as a function of energy and polar angle \(\theta\) in the \(yz\) plane, with the positive \(z\) axis lying along \(\theta=0^{\circ}\) and the positive \(y\) axis along \(\theta = 90^{\circ}\). 
The measured distribution consists of two sets of three maxima, located either side of the \(z\) (vertical) axis. The origin of these features may be visualized using Fig.\;\ref{fig:3dplot}, with a maximum arising from each of the six spiral arms. The innermost feature, close to \(E=0.25\) eV, is rather weak, since the \(y\) axis cuts through a spiral tail at this energy, as can be seen in Fig.\;\ref{exp20fscomp}(a). The strongest feature appears at around \(E=0.35\) eV, and maintains significant magnitude over a \(90^{\circ}\) sector before dying away. The same is true of the outermost feature at around \(E=0.45\) eV. Note that in the perturbative limit, the \(yz\) plane distribution may be obtained from Eq.\;\eqref{pertex}, and takes the form
\begin{equation}
    P(E,\theta,\phi=\pm\pi/2) = g(E)\sin^6\theta \left[1+\cos E\tau\right],
    \label{pyz}
\end{equation}
which again implies a strong time-delay dependence within restricted angular ranges close to \(\theta = \pm\pi/2\).

The distributions calculated using RMT (Figs.\;\ref{exp20fsyz}(b)-(d)) also present these features, but their respective magnitudes and energy locations display the expected sensitivity to time delay between the pulses.  At \(\tau=40\) fs, the outermost feature is rather weak, and all of the peaks appear at marginally higher energies than those of the measurement. At \(\tau=38\) fs, the peaks shift to a higher energy still, with the outermost feature now appearing very weakly at around \(E=0.5\) eV. However, at \(\tau=42\) fs, all features shift towards lower energies, and the relative magnitudes show a noticeably improved agreement with experiment. The time-delay dependence of both the energy shift and relative prominence of each of the features may be trivially related to the time-delay dependence in the spiral orientation, as can be visualized using Figs.\;\ref{fig:3dplot} and \ref{exp20fscomp}. The variation in the location of the maxima may also be inferred from the cosine term in Eq.\;\eqref{pyz}: as \(\tau\) increases, the energy at which a maximum occurs decreases. Hence, in this case, since the dominant features in the calculation using \(\tau = 38\) fs lie at a higher energy than those in experiment, the time delay should be progressively increased to improve the level of agreement.

Although the distributions in Fig.\;\ref{exp20fsyz} display less structure than those of Fig.\;\ref{exp20fscomp}, they are potentially more useful for the purpose of time-delay calibration. The main features in the \(yz\) plane are clearly well localized in angle, due to the helpful \(\sin^6\theta\) dependence, and their intensity varies strongly as a function of energy and time delay, thereby providing three strong categories of correspondence with measurement. In the \(xy\) plane, the distribution spans the full angular range, oscillates strongly in angle and energy due to the \(\cos(6\phi+E\tau)\) term, and rotates as a function of time delay, meaning that comparison with measurement involves a detailed assessment of the angular extent of the spiral arms, and the yield variation along them. Therefore, analysis of the distribution in planes perpendicular to the laser polarization provides an effective means for calibration of measured time delays between two pulses.

%

To explore this application further, we show in Fig.\;\ref{exp20fsxz} the projection of the distribution on the \(xz\) plane. Six maxima are again expected in this plane, though their relative prominence may differ from those in the \(yz\)-plane projection. This is indeed the case, as the measured distribution of Fig.\;\ref{exp20fsxz}(a) displays two sets of two, almost commensurate maxima, and a third, much weaker feature on either side of the \(z\) (vertical) axis. The weakness of the third feature (at around 0.2 eV) may be expected given the distribution measured in the \(xy\) plane (Fig.\;\ref{exp20fscomp}(a)), in which the innermost spiral arm tails off as it approaches the \(x\) axis (\(\phi=0^{\circ},180^{\circ}\)). In this case, the calculation using a 42-fs delay agrees very well with the measurement, with the stronger feature, centered close to \(E=0.4\) eV, sweeping out roughly a \(90^{\circ}\) sector, and the feature around \(E=0.3\) eV sweeping round a slightly smaller angle. 

We note that the difference between the distributions in the \(xz\) and \(yz\) planes may be explained using the perturbative analysis embodied by Eq.\;\eqref{pertex}. In contrast to the predicted distribution in the \(yz\) plane, given by Eq.\;\eqref{pyz}, in the \(xz\) plane the distribution is given by
\begin{equation}
    P(E,\theta,\phi=0,\pi) = g(E)\sin^6\theta \left[1-\cos E\tau\right].
    \label{pxz}
\end{equation}
Thus, in the perturbative limit, the energies at which maxima are observed in one plane will be those at which minima are observed in the other. This phase difference is clearly manifest in both the measured and calculated distributions (at all time delays). Eq.\;\eqref{pxz} indicates that the trend seen in Fig.\;\ref{exp20fsyz}, where the dominant features shift to lower energies as time delay increases, should be preserved in the \(xz\) plane distributions. Such a trend is indeed apparent in Fig.\;\ref{exp20fsxz}, with the dominant features progressively shifting into line with those of the measurement as time delay is increased.

\begin{figure}[t]
    \centering
    \includegraphics[width=\columnwidth]{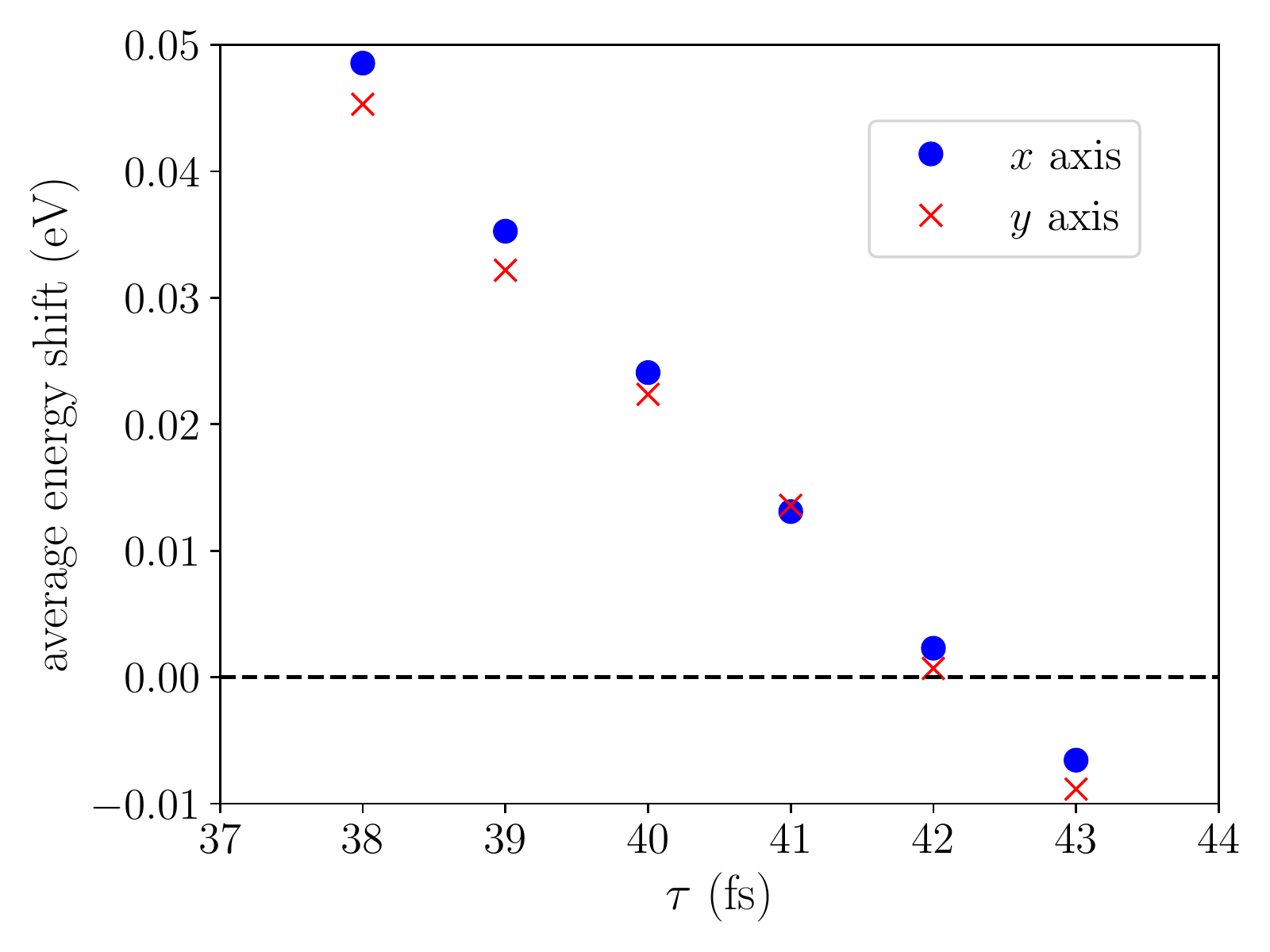}
    \caption{Average shift in energy of the maxima along the \(x\) and \(y\) directions relative to experimental measurement \cite{pengel2017a}, as a function of the time delay \(\tau\) between the laser pulses.}
    \label{devs}
\end{figure}

In fact, such shifts can play a critical role in calibrating the time delay between the laser pulses. Given the form of Eqs.\;\eqref{pyz} and \eqref{pxz}, the time-delay dependence may be effectively isolated by fixing \(\theta\), and considering slices of the distribution in given directions. To characterize these distributions, we consider the energy dependence in the \(x\) and \(y\) directions, both of which consist of a set of three maxima (see Figs.\;\ref{exp20fsyz} and \ref{exp20fsxz}). At each time delay, we compare the energy locations of the maxima appearing in our calculations with those observed in experiment. We then calculate the average deviation in these energies from the experimental values. Figure \ref{devs} shows the average energy shift relative to experiment of the maxima located along the \(x\) and \(y\) axes as a function of time delay (including calculations with \(\tau = 39, 41, 43\) fs not shown in Figs.\;\ref{exp20fscomp}, \ref{exp20fsyz} and \ref{exp20fsxz}). Over this range of delays, the relationship is approximately linear in both cases. The estimated time delay is therefore the zero of the average shift, and can be found simply through linear interpolation. The energy shifts in the \(x\) direction yield an estimated time delay of 42.27 fs. Using the data in the \(y\)-direction, the estimate is 42.14 fs. With this method, we are able to achieve a significant improvement on the experimental uncertainty, which in these cases is around 2 fs.

Finally, we note that the potential for time-delay calibration could be further enhanced by additional photon absorptions. If ionization is achieved through \(n\) photon absorptions from each pulse, the \(xy\)-plane distribution will display a \(\cos(2n\phi + E\tau)\) dependence in the perturbative limit, yielding a \(2n\)-arm spiral. In the \(yz\) or \(xz\) planes, the distribution will contain a \(\sin^{2n}\theta\) dependence, providing increasing localization of the main features as \(n\) increases, and the \(\cos E\tau\) term will be maintained for all \(n\). Therefore, as \(n\) increases, the \(xy\)-plane distributions become ever more complex, while those in perpendicular planes reduce in complexity. These distributions should generally offer the best opportunities for calibration of measured time delays, with quality likely to increase with the number of photons absorbed.

\section{Conclusion}
In this work, we have demonstrated the accuracy of the RMT approach for arbitrarily polarized laser fields through comparison with recent experimental tomographic measurements \cite{pengel2017a}. Specifically, we have investigated the multiphoton ionization of potassium by a pair of counter-rotating circularly polarized pulses, and calculated the full-dimensional energy-space probability distribution using solutions of the multielectron time-dependent Schr\"{o}dinger equation. Calculations have been performed for a range of laser intensities and time delays between the two pulses, covering the experimental uncertainty in these quantities.

A clear comparison with the measured data presented in Ref.\;\cite{pengel2017a} is achieved by examining planar slices of the measured and calculated distributions. We compare the respective results in the laser polarization plane, and in planes perpendicular to it. We find that our calculations agree well with the measurement over the experimental range of laser intensities and time delays, with particularly strong agreement achieved for a time delay of 42 fs between the pulses. The characteristics of the calculated distributions might therefore be used as a means of calibrating measured time delays between counter-rotating circularly polarized pulses.

The physical character of the distribution naturally varies strongly from plane to plane. In the laser polarization plane, the well-known spiral features emerge, whereas perpendicular to the laser polarization, a multilobe structure appears. Our calculations show strong agreement with each of these features, indicating an accurate, full-dimensional characterization of the dynamics. Further investigation of the time-delay dependence of the calculated distributions may enable effective calibration of the measured time delay, allowing it to be fine tuned within the experimental error estimate.

\section*{Acknowledgments}
We thank Dr Stefanie Kerbstadt and Professor Matthias Wollenhaupt for providing their experimental data in numerical form. The RMT data presented in this work may be accessed using Ref.\;\cite{pure}. We thank Zden\v{e}k Ma\v{s}\'{i}n and Jimena Gorfinkiel for collaboration in developing and maintaining the RMT code. The RMT code is part of the UK-AMOR suite, and can be obtained for free from Ref.\;\cite{repo}. 
This work benefited from computational support by CoSeC, the Computational Science Centre for Research Communities, through CCPQ. DDAC acknowledges financial support from the UK Engineering and Physical Sciences Research Council (EPSRC). The authors acknowledge funding from the EPSRC under grants EP/P022146/1, EP/P013953/1 and EP/R029342/1. The calculations reported in this work relied on the ARCHER UK National Supercomputing Service (\href{www.archer.ac.uk}{www.archer.ac.uk}) for which access was obtained via the UK-AMOR consortium funded by EPSRC.

\appendix

\section{Photoelectron momentum distribution}
\label{momdisapp}
In atomic RMT, the outer-region wave function is expanded in channels, $p$, labeled by the total angular momentum, \(L\), and its projection, \(M_L\), the residual-ion state with total angular momentum, \(L_T\), and its projection, \(M_{T}\), as well as the angular momentum, \(l\), of the photoelectron with projection, \(m\).
The photoelectron momentum distribution, $\rho({\bf k}_{N+1})$, is given by
\begin{widetext}
\beq
    \rho({\bf k}_{N+1}) = \sum_{L_T M_{T}}
    \left\vert 
        \sum_{l m LM_L}  (L_T M_{T} l m | L M_L) i^l G_{L_T M_T l m LM_L} (k_{N+1}){\cal Y}_{l m}(\theta_{N+1},\phi_{N+1})
    \right\vert^2 ,
\eeq
\end{widetext}
where \((L_T M_{T} l m | L M_L)\) is a Clebsch-Gordan coefficient, \({\cal Y}_{lm}\) is a complex spherical harmonic conforming to the Fano-Racah phase convention \cite{fano1959}, and \(G_{L_T M_T l m LM} (k_{N+1})\) is the Fourier transform of the position-space photoelectron wavefunction \(F_{L_T M_T l m LM} (r_{N+1})\) (integrating from a minimum radial distance \(r_0\) to the maximum radial extent of the outer region \(r_{\rm max}\)),
\beq
    G_{p} (k_{N+1}) = \int_{r_0}^{r_{\rm max}} e^{-ik_{N+1}r_{N+1}} F_{p} (r_{N+1})\ dr_{N+1} .
\eeq
In this work, we have chosen \(r_0\) to be the distance at which the outer-region begins, so that \(r_0 = 50\textrm{ a.u.}\), and \(r_{\rm max} = 4750\textrm{ a.u.}\). We have verified that all contributing photoelectron wavepackets have reached far beyond the inner-region radius, \(r_0\). 
\bibliography{mybib}

\begin{thebibliography}{68}%
\makeatletter
\providecommand \@ifxundefined [1]{%
 \@ifx{#1\undefined}
}%
\providecommand \@ifnum [1]{%
 \ifnum #1\expandafter \@firstoftwo
 \else \expandafter \@secondoftwo
 \fi
}%
\providecommand \@ifx [1]{%
 \ifx #1\expandafter \@firstoftwo
 \else \expandafter \@secondoftwo
 \fi
}%
\providecommand \natexlab [1]{#1}%
\providecommand \enquote  [1]{``#1''}%
\providecommand \bibnamefont  [1]{#1}%
\providecommand \bibfnamefont [1]{#1}%
\providecommand \citenamefont [1]{#1}%
\providecommand \href@noop [0]{\@secondoftwo}%
\providecommand \href [0]{\begingroup \@sanitize@url \@href}%
\providecommand \@href[1]{\@@startlink{#1}\@@href}%
\providecommand \@@href[1]{\endgroup#1\@@endlink}%
\providecommand \@sanitize@url [0]{\catcode `\\12\catcode `\$12\catcode
  `\&12\catcode `\#12\catcode `\^12\catcode `\_12\catcode `\%12\relax}%
\providecommand \@@startlink[1]{}%
\providecommand \@@endlink[0]{}%
\providecommand \url  [0]{\begingroup\@sanitize@url \@url }%
\providecommand \@url [1]{\endgroup\@href {#1}{\urlprefix }}%
\providecommand \urlprefix  [0]{URL }%
\providecommand \Eprint [0]{\href }%
\providecommand \doibase [0]{http://dx.doi.org/}%
\providecommand \selectlanguage [0]{\@gobble}%
\providecommand \bibinfo  [0]{\@secondoftwo}%
\providecommand \bibfield  [0]{\@secondoftwo}%
\providecommand \translation [1]{[#1]}%
\providecommand \BibitemOpen [0]{}%
\providecommand \bibitemStop [0]{}%
\providecommand \bibitemNoStop [0]{.\EOS\space}%
\providecommand \EOS [0]{\spacefactor3000\relax}%
\providecommand \BibitemShut  [1]{\csname bibitem#1\endcsname}%
\let\auto@bib@innerbib\@empty
\bibitem [{\citenamefont {Silberberg}(2009)}]{silberberg2009}%
  \BibitemOpen
  \bibfield  {author} {\bibinfo {author} {\bibfnamefont {Y.}~\bibnamefont
  {Silberberg}},\ }\href {\doibase 10.1146/annurev.physchem.040808.090427}
  {\bibfield  {journal} {\bibinfo  {journal} {Annual Review of Physical
  Chemistry}\ }\textbf {\bibinfo {volume} {60}},\ \bibinfo {pages} {277}
  (\bibinfo {year} {2009})},\ \bibinfo {note} {pMID: 18999997}\BibitemShut
  {NoStop}%
\bibitem [{\citenamefont {Misawa}(2016)}]{misawa2016}%
  \BibitemOpen
  \bibfield  {author} {\bibinfo {author} {\bibfnamefont {K.}~\bibnamefont
  {Misawa}},\ }\href {\doibase 10.1080/23746149.2016.1221327} {\bibfield
  {journal} {\bibinfo  {journal} {Advances in Physics: X}\ }\textbf {\bibinfo
  {volume} {1}},\ \bibinfo {pages} {544} (\bibinfo {year} {2016})}\BibitemShut
  {NoStop}%
\bibitem [{\citenamefont {Wollenhaupt}\ \emph {et~al.}(2016)\citenamefont
  {Wollenhaupt}, \citenamefont {Bayer},\ and\ \citenamefont
  {Baumert}}]{wollenhaupt2016}%
  \BibitemOpen
  \bibfield  {author} {\bibinfo {author} {\bibfnamefont {M.}~\bibnamefont
  {Wollenhaupt}}, \bibinfo {author} {\bibfnamefont {T.}~\bibnamefont {Bayer}},
  \ and\ \bibinfo {author} {\bibfnamefont {T.}~\bibnamefont {Baumert}},\
  }\href@noop {} {\emph {\bibinfo {title} {Control of ultrafast electron
  dynamics with shaped femtosecond laser pulses: From atoms to solids, in
  Ultrafast Dynamics Driven by Intense Light Pulses: From Atoms to Solids, from
  Lasers to Intense X-rays}}}\ (\bibinfo  {publisher} {Springer International,
  Cham},\ \bibinfo {year} {2016})\BibitemShut {NoStop}%
\bibitem [{\citenamefont {Brixner}\ and\ \citenamefont
  {Gerber}(2001)}]{brixner2001}%
  \BibitemOpen
  \bibfield  {author} {\bibinfo {author} {\bibfnamefont {T.}~\bibnamefont
  {Brixner}}\ and\ \bibinfo {author} {\bibfnamefont {G.}~\bibnamefont
  {Gerber}},\ }\href {\doibase 10.1364/OL.26.000557} {\bibfield  {journal}
  {\bibinfo  {journal} {Opt. Lett.}\ }\textbf {\bibinfo {volume} {26}},\
  \bibinfo {pages} {557} (\bibinfo {year} {2001})}\BibitemShut {NoStop}%
\bibitem [{\citenamefont {Brixner}\ \emph {et~al.}(2004)\citenamefont
  {Brixner}, \citenamefont {Krampert}, \citenamefont {Pfeifer}, \citenamefont
  {Selle}, \citenamefont {Gerber}, \citenamefont {Wollenhaupt}, \citenamefont
  {Graefe}, \citenamefont {Horn}, \citenamefont {Liese},\ and\ \citenamefont
  {Baumert}}]{brixner2004}%
  \BibitemOpen
  \bibfield  {author} {\bibinfo {author} {\bibfnamefont {T.}~\bibnamefont
  {Brixner}}, \bibinfo {author} {\bibfnamefont {G.}~\bibnamefont {Krampert}},
  \bibinfo {author} {\bibfnamefont {T.}~\bibnamefont {Pfeifer}}, \bibinfo
  {author} {\bibfnamefont {R.}~\bibnamefont {Selle}}, \bibinfo {author}
  {\bibfnamefont {G.}~\bibnamefont {Gerber}}, \bibinfo {author} {\bibfnamefont
  {M.}~\bibnamefont {Wollenhaupt}}, \bibinfo {author} {\bibfnamefont
  {O.}~\bibnamefont {Graefe}}, \bibinfo {author} {\bibfnamefont
  {C.}~\bibnamefont {Horn}}, \bibinfo {author} {\bibfnamefont {D.}~\bibnamefont
  {Liese}}, \ and\ \bibinfo {author} {\bibfnamefont {T.}~\bibnamefont
  {Baumert}},\ }\href {\doibase 10.1103/PhysRevLett.92.208301} {\bibfield
  {journal} {\bibinfo  {journal} {Phys. Rev. Lett.}\ }\textbf {\bibinfo
  {volume} {92}},\ \bibinfo {pages} {208301} (\bibinfo {year}
  {2004})}\BibitemShut {NoStop}%
\bibitem [{\citenamefont {Suzuki}\ \emph {et~al.}(2004)\citenamefont {Suzuki},
  \citenamefont {Minemoto}, \citenamefont {Kanai},\ and\ \citenamefont
  {Sakai}}]{suzuki2004}%
  \BibitemOpen
  \bibfield  {author} {\bibinfo {author} {\bibfnamefont {T.}~\bibnamefont
  {Suzuki}}, \bibinfo {author} {\bibfnamefont {S.}~\bibnamefont {Minemoto}},
  \bibinfo {author} {\bibfnamefont {T.}~\bibnamefont {Kanai}}, \ and\ \bibinfo
  {author} {\bibfnamefont {H.}~\bibnamefont {Sakai}},\ }\href {\doibase
  10.1103/PhysRevLett.92.133005} {\bibfield  {journal} {\bibinfo  {journal}
  {Phys. Rev. Lett.}\ }\textbf {\bibinfo {volume} {92}},\ \bibinfo {pages}
  {133005} (\bibinfo {year} {2004})}\BibitemShut {NoStop}%
\bibitem [{\citenamefont {Dudovich}\ \emph {et~al.}(2004)\citenamefont
  {Dudovich}, \citenamefont {Oron},\ and\ \citenamefont
  {Silberberg}}]{dudovich2004}%
  \BibitemOpen
  \bibfield  {author} {\bibinfo {author} {\bibfnamefont {N.}~\bibnamefont
  {Dudovich}}, \bibinfo {author} {\bibfnamefont {D.}~\bibnamefont {Oron}}, \
  and\ \bibinfo {author} {\bibfnamefont {Y.}~\bibnamefont {Silberberg}},\
  }\href {\doibase 10.1103/PhysRevLett.92.103003} {\bibfield  {journal}
  {\bibinfo  {journal} {Phys. Rev. Lett.}\ }\textbf {\bibinfo {volume} {92}},\
  \bibinfo {pages} {103003} (\bibinfo {year} {2004})}\BibitemShut {NoStop}%
\bibitem [{\citenamefont {Kerbstadt}\ \emph
  {et~al.}(2017{\natexlab{a}})\citenamefont {Kerbstadt}, \citenamefont
  {Timmer}, \citenamefont {Englert}, \citenamefont {Bayer},\ and\ \citenamefont
  {Wollenhaupt}}]{kerbstadt2017a}%
  \BibitemOpen
  \bibfield  {author} {\bibinfo {author} {\bibfnamefont {S.}~\bibnamefont
  {Kerbstadt}}, \bibinfo {author} {\bibfnamefont {D.}~\bibnamefont {Timmer}},
  \bibinfo {author} {\bibfnamefont {L.}~\bibnamefont {Englert}}, \bibinfo
  {author} {\bibfnamefont {T.}~\bibnamefont {Bayer}}, \ and\ \bibinfo {author}
  {\bibfnamefont {M.}~\bibnamefont {Wollenhaupt}},\ }\href {\doibase
  10.1364/OE.25.012518} {\bibfield  {journal} {\bibinfo  {journal} {Opt.
  Express}\ }\textbf {\bibinfo {volume} {25}},\ \bibinfo {pages} {12518}
  (\bibinfo {year} {2017}{\natexlab{a}})}\BibitemShut {NoStop}%
\bibitem [{\citenamefont {Kerbstadt}\ \emph
  {et~al.}(2017{\natexlab{b}})\citenamefont {Kerbstadt}, \citenamefont
  {Englert}, \citenamefont {Bayer},\ and\ \citenamefont
  {Wollenhaupt}}]{kerbstadt2017b}%
  \BibitemOpen
  \bibfield  {author} {\bibinfo {author} {\bibfnamefont {S.}~\bibnamefont
  {Kerbstadt}}, \bibinfo {author} {\bibfnamefont {L.}~\bibnamefont {Englert}},
  \bibinfo {author} {\bibfnamefont {T.}~\bibnamefont {Bayer}}, \ and\ \bibinfo
  {author} {\bibfnamefont {M.}~\bibnamefont {Wollenhaupt}},\ }\href {\doibase
  10.1080/09500340.2016.1271151} {\bibfield  {journal} {\bibinfo  {journal}
  {Journal of Modern Optics}\ }\textbf {\bibinfo {volume} {64}},\ \bibinfo
  {pages} {1010} (\bibinfo {year} {2017}{\natexlab{b}})}\BibitemShut {NoStop}%
\bibitem [{\citenamefont {Pengel}\ \emph
  {et~al.}(2017{\natexlab{a}})\citenamefont {Pengel}, \citenamefont
  {Kerbstadt}, \citenamefont {Johannmeyer}, \citenamefont {Englert},
  \citenamefont {Bayer},\ and\ \citenamefont {Wollenhaupt}}]{pengel2017}%
  \BibitemOpen
  \bibfield  {author} {\bibinfo {author} {\bibfnamefont {D.}~\bibnamefont
  {Pengel}}, \bibinfo {author} {\bibfnamefont {S.}~\bibnamefont {Kerbstadt}},
  \bibinfo {author} {\bibfnamefont {D.}~\bibnamefont {Johannmeyer}}, \bibinfo
  {author} {\bibfnamefont {L.}~\bibnamefont {Englert}}, \bibinfo {author}
  {\bibfnamefont {T.}~\bibnamefont {Bayer}}, \ and\ \bibinfo {author}
  {\bibfnamefont {M.}~\bibnamefont {Wollenhaupt}},\ }\href {\doibase
  10.1103/PhysRevLett.118.053003} {\bibfield  {journal} {\bibinfo  {journal}
  {Phys. Rev. Lett.}\ }\textbf {\bibinfo {volume} {118}},\ \bibinfo {pages}
  {053003} (\bibinfo {year} {2017}{\natexlab{a}})}\BibitemShut {NoStop}%
\bibitem [{\citenamefont {Pengel}\ \emph
  {et~al.}(2017{\natexlab{b}})\citenamefont {Pengel}, \citenamefont
  {Kerbstadt}, \citenamefont {Englert}, \citenamefont {Bayer},\ and\
  \citenamefont {Wollenhaupt}}]{pengel2017a}%
  \BibitemOpen
  \bibfield  {author} {\bibinfo {author} {\bibfnamefont {D.}~\bibnamefont
  {Pengel}}, \bibinfo {author} {\bibfnamefont {S.}~\bibnamefont {Kerbstadt}},
  \bibinfo {author} {\bibfnamefont {L.}~\bibnamefont {Englert}}, \bibinfo
  {author} {\bibfnamefont {T.}~\bibnamefont {Bayer}}, \ and\ \bibinfo {author}
  {\bibfnamefont {M.}~\bibnamefont {Wollenhaupt}},\ }\href {\doibase
  10.1103/PhysRevA.96.043426} {\bibfield  {journal} {\bibinfo  {journal} {Phys.
  Rev. A}\ }\textbf {\bibinfo {volume} {96}},\ \bibinfo {pages} {043426}
  (\bibinfo {year} {2017}{\natexlab{b}})}\BibitemShut {NoStop}%
\bibitem [{\citenamefont {Kerbstadt}\ \emph {et~al.}(2019)\citenamefont
  {Kerbstadt}, \citenamefont {Eickhoff}, \citenamefont {Bayer},\ and\
  \citenamefont {Wollenhaupt}}]{kerbstadt2019}%
  \BibitemOpen
  \bibfield  {author} {\bibinfo {author} {\bibfnamefont {S.}~\bibnamefont
  {Kerbstadt}}, \bibinfo {author} {\bibfnamefont {K.}~\bibnamefont {Eickhoff}},
  \bibinfo {author} {\bibfnamefont {T.}~\bibnamefont {Bayer}}, \ and\ \bibinfo
  {author} {\bibfnamefont {M.}~\bibnamefont {Wollenhaupt}},\ }\href {\doibase
  10.1038/s41467-019-08601-7} {\bibfield  {journal} {\bibinfo  {journal}
  {Nature Communications}\ }\textbf {\bibinfo {volume} {10}},\ \bibinfo {pages}
  {658} (\bibinfo {year} {2019})}\BibitemShut {NoStop}%
\bibitem [{\citenamefont {Zuo}\ and\ \citenamefont {Bandrauk}(1995)}]{zuo1995}%
  \BibitemOpen
  \bibfield  {author} {\bibinfo {author} {\bibfnamefont {T.}~\bibnamefont
  {Zuo}}\ and\ \bibinfo {author} {\bibfnamefont {A.~D.}\ \bibnamefont
  {Bandrauk}},\ }\href {\doibase 10.1142/S0218863595000227} {\bibfield
  {journal} {\bibinfo  {journal} {Journal of Nonlinear Optical Physics \&
  Materials}\ }\textbf {\bibinfo {volume} {04}},\ \bibinfo {pages} {533}
  (\bibinfo {year} {1995})}\BibitemShut {NoStop}%
\bibitem [{\citenamefont {Long}\ \emph {et~al.}(1995)\citenamefont {Long},
  \citenamefont {Becker},\ and\ \citenamefont {McIver}}]{long1995}%
  \BibitemOpen
  \bibfield  {author} {\bibinfo {author} {\bibfnamefont {S.}~\bibnamefont
  {Long}}, \bibinfo {author} {\bibfnamefont {W.}~\bibnamefont {Becker}}, \ and\
  \bibinfo {author} {\bibfnamefont {J.~K.}\ \bibnamefont {McIver}},\ }\href
  {\doibase 10.1103/PhysRevA.52.2262} {\bibfield  {journal} {\bibinfo
  {journal} {Phys. Rev. A}\ }\textbf {\bibinfo {volume} {52}},\ \bibinfo
  {pages} {2262} (\bibinfo {year} {1995})}\BibitemShut {NoStop}%
\bibitem [{\citenamefont {Milo\ifmmode \check{s}\else
  \v{s}\fi{}evi\ifmmode~\acute{c}\else \'{c}\fi{}}\ \emph
  {et~al.}(2000)\citenamefont {Milo\ifmmode \check{s}\else
  \v{s}\fi{}evi\ifmmode~\acute{c}\else \'{c}\fi{}}, \citenamefont {Becker},\
  and\ \citenamefont {Kopold}}]{milosevic2000}%
  \BibitemOpen
  \bibfield  {author} {\bibinfo {author} {\bibfnamefont {D.~B.}\ \bibnamefont
  {Milo\ifmmode \check{s}\else \v{s}\fi{}evi\ifmmode~\acute{c}\else
  \'{c}\fi{}}}, \bibinfo {author} {\bibfnamefont {W.}~\bibnamefont {Becker}}, \
  and\ \bibinfo {author} {\bibfnamefont {R.}~\bibnamefont {Kopold}},\ }\href
  {\doibase 10.1103/PhysRevA.61.063403} {\bibfield  {journal} {\bibinfo
  {journal} {Phys. Rev. A}\ }\textbf {\bibinfo {volume} {61}},\ \bibinfo
  {pages} {063403} (\bibinfo {year} {2000})}\BibitemShut {NoStop}%
\bibitem [{\citenamefont {Fleischer}\ \emph {et~al.}(2014)\citenamefont
  {Fleischer}, \citenamefont {Kfir}, \citenamefont {Diskin}, \citenamefont
  {Sidorenko},\ and\ \citenamefont {Cohen}}]{fleischer2014}%
  \BibitemOpen
  \bibfield  {author} {\bibinfo {author} {\bibfnamefont {A.}~\bibnamefont
  {Fleischer}}, \bibinfo {author} {\bibfnamefont {O.}~\bibnamefont {Kfir}},
  \bibinfo {author} {\bibfnamefont {T.}~\bibnamefont {Diskin}}, \bibinfo
  {author} {\bibfnamefont {P.}~\bibnamefont {Sidorenko}}, \ and\ \bibinfo
  {author} {\bibfnamefont {O.}~\bibnamefont {Cohen}},\ }\href
  {https://doi.org/10.1038/nphoton.2014.108} {\bibfield  {journal} {\bibinfo
  {journal} {Nature Photonics}\ }\textbf {\bibinfo {volume} {8}},\ \bibinfo
  {pages} {543} (\bibinfo {year} {2014})}\BibitemShut {NoStop}%
\bibitem [{\citenamefont {Milo\ifmmode \check{s}\else
  \v{s}\fi{}evi\ifmmode~\acute{c}\else \'{c}\fi{}}(2015)}]{milosevic2015}%
  \BibitemOpen
  \bibfield  {author} {\bibinfo {author} {\bibfnamefont {D.~B.}\ \bibnamefont
  {Milo\ifmmode \check{s}\else \v{s}\fi{}evi\ifmmode~\acute{c}\else
  \'{c}\fi{}}},\ }\href {\doibase 10.1103/PhysRevA.92.043827} {\bibfield
  {journal} {\bibinfo  {journal} {Phys. Rev. A}\ }\textbf {\bibinfo {volume}
  {92}},\ \bibinfo {pages} {043827} (\bibinfo {year} {2015})}\BibitemShut
  {NoStop}%
\bibitem [{\citenamefont {Kfir}\ \emph {et~al.}(2014)\citenamefont {Kfir},
  \citenamefont {Grychtol}, \citenamefont {Turgut}, \citenamefont {Knut},
  \citenamefont {Zusin}, \citenamefont {Popmintchev}, \citenamefont
  {Popmintchev}, \citenamefont {Nembach}, \citenamefont {Shaw}, \citenamefont
  {Fleischer}, \citenamefont {Kapteyn}, \citenamefont {Murnane},\ and\
  \citenamefont {Cohen}}]{kfir2015}%
  \BibitemOpen
  \bibfield  {author} {\bibinfo {author} {\bibfnamefont {O.}~\bibnamefont
  {Kfir}}, \bibinfo {author} {\bibfnamefont {P.}~\bibnamefont {Grychtol}},
  \bibinfo {author} {\bibfnamefont {E.}~\bibnamefont {Turgut}}, \bibinfo
  {author} {\bibfnamefont {R.}~\bibnamefont {Knut}}, \bibinfo {author}
  {\bibfnamefont {D.}~\bibnamefont {Zusin}}, \bibinfo {author} {\bibfnamefont
  {D.}~\bibnamefont {Popmintchev}}, \bibinfo {author} {\bibfnamefont
  {T.}~\bibnamefont {Popmintchev}}, \bibinfo {author} {\bibfnamefont
  {H.}~\bibnamefont {Nembach}}, \bibinfo {author} {\bibfnamefont {J.~M.}\
  \bibnamefont {Shaw}}, \bibinfo {author} {\bibfnamefont {A.}~\bibnamefont
  {Fleischer}}, \bibinfo {author} {\bibfnamefont {H.}~\bibnamefont {Kapteyn}},
  \bibinfo {author} {\bibfnamefont {M.}~\bibnamefont {Murnane}}, \ and\
  \bibinfo {author} {\bibfnamefont {O.}~\bibnamefont {Cohen}},\ }\href
  {https://doi.org/10.1038/nphoton.2014.293} {\bibfield  {journal} {\bibinfo
  {journal} {Nature Photonics}\ }\textbf {\bibinfo {volume} {9}},\ \bibinfo
  {pages} {99} (\bibinfo {year} {2014})}\BibitemShut {NoStop}%
\bibitem [{\citenamefont {Medi\ifmmode~\check{s}\else \v{s}\fi{}auskas}\ \emph
  {et~al.}(2015)\citenamefont {Medi\ifmmode~\check{s}\else \v{s}\fi{}auskas},
  \citenamefont {Wragg}, \citenamefont {van~der Hart},\ and\ \citenamefont
  {Ivanov}}]{lukas2015}%
  \BibitemOpen
  \bibfield  {author} {\bibinfo {author} {\bibfnamefont {L.}~\bibnamefont
  {Medi\ifmmode~\check{s}\else \v{s}\fi{}auskas}}, \bibinfo {author}
  {\bibfnamefont {J.}~\bibnamefont {Wragg}}, \bibinfo {author} {\bibfnamefont
  {H.}~\bibnamefont {van~der Hart}}, \ and\ \bibinfo {author} {\bibfnamefont
  {M.~Y.}\ \bibnamefont {Ivanov}},\ }\href {\doibase
  10.1103/PhysRevLett.115.153001} {\bibfield  {journal} {\bibinfo  {journal}
  {Phys. Rev. Lett.}\ }\textbf {\bibinfo {volume} {115}},\ \bibinfo {pages}
  {153001} (\bibinfo {year} {2015})}\BibitemShut {NoStop}%
\bibitem [{\citenamefont {Dorney}\ \emph {et~al.}(2017)\citenamefont {Dorney},
  \citenamefont {Ellis}, \citenamefont {Hern\'andez-Garc\'{\i}a}, \citenamefont
  {Hickstein}, \citenamefont {Mancuso}, \citenamefont {Brooks}, \citenamefont
  {Fan}, \citenamefont {Fan}, \citenamefont {Zusin}, \citenamefont {Gentry},
  \citenamefont {Grychtol}, \citenamefont {Kapteyn},\ and\ \citenamefont
  {Murnane}}]{dorney2017}%
  \BibitemOpen
  \bibfield  {author} {\bibinfo {author} {\bibfnamefont {K.~M.}\ \bibnamefont
  {Dorney}}, \bibinfo {author} {\bibfnamefont {J.~L.}\ \bibnamefont {Ellis}},
  \bibinfo {author} {\bibfnamefont {C.}~\bibnamefont
  {Hern\'andez-Garc\'{\i}a}}, \bibinfo {author} {\bibfnamefont {D.~D.}\
  \bibnamefont {Hickstein}}, \bibinfo {author} {\bibfnamefont {C.~A.}\
  \bibnamefont {Mancuso}}, \bibinfo {author} {\bibfnamefont {N.}~\bibnamefont
  {Brooks}}, \bibinfo {author} {\bibfnamefont {T.}~\bibnamefont {Fan}},
  \bibinfo {author} {\bibfnamefont {G.}~\bibnamefont {Fan}}, \bibinfo {author}
  {\bibfnamefont {D.}~\bibnamefont {Zusin}}, \bibinfo {author} {\bibfnamefont
  {C.}~\bibnamefont {Gentry}}, \bibinfo {author} {\bibfnamefont
  {P.}~\bibnamefont {Grychtol}}, \bibinfo {author} {\bibfnamefont {H.~C.}\
  \bibnamefont {Kapteyn}}, \ and\ \bibinfo {author} {\bibfnamefont {M.~M.}\
  \bibnamefont {Murnane}},\ }\href {\doibase 10.1103/PhysRevLett.119.063201}
  {\bibfield  {journal} {\bibinfo  {journal} {Phys. Rev. Lett.}\ }\textbf
  {\bibinfo {volume} {119}},\ \bibinfo {pages} {063201} (\bibinfo {year}
  {2017})}\BibitemShut {NoStop}%
\bibitem [{\citenamefont {Wollenhaupt}\ \emph {et~al.}(2009)\citenamefont
  {Wollenhaupt}, \citenamefont {Krug}, \citenamefont {K{\"o}hler},
  \citenamefont {Bayer}, \citenamefont {Sarpe-Tudoran},\ and\ \citenamefont
  {Baumert}}]{wollenhaupt2009}%
  \BibitemOpen
  \bibfield  {author} {\bibinfo {author} {\bibfnamefont {M.}~\bibnamefont
  {Wollenhaupt}}, \bibinfo {author} {\bibfnamefont {M.}~\bibnamefont {Krug}},
  \bibinfo {author} {\bibfnamefont {J.}~\bibnamefont {K{\"o}hler}}, \bibinfo
  {author} {\bibfnamefont {T.}~\bibnamefont {Bayer}}, \bibinfo {author}
  {\bibfnamefont {C.}~\bibnamefont {Sarpe-Tudoran}}, \ and\ \bibinfo {author}
  {\bibfnamefont {T.}~\bibnamefont {Baumert}},\ }\href {\doibase
  10.1007/s00340-009-3513-0} {\bibfield  {journal} {\bibinfo  {journal}
  {Applied Physics B}\ }\textbf {\bibinfo {volume} {95}},\ \bibinfo {pages}
  {647} (\bibinfo {year} {2009})}\BibitemShut {NoStop}%
\bibitem [{\citenamefont {Smeenk}\ \emph {et~al.}(2009)\citenamefont {Smeenk},
  \citenamefont {Arissian}, \citenamefont {Staudte}, \citenamefont
  {Villeneuve},\ and\ \citenamefont {Corkum}}]{smeenk2009}%
  \BibitemOpen
  \bibfield  {author} {\bibinfo {author} {\bibfnamefont {C.}~\bibnamefont
  {Smeenk}}, \bibinfo {author} {\bibfnamefont {L.}~\bibnamefont {Arissian}},
  \bibinfo {author} {\bibfnamefont {A.}~\bibnamefont {Staudte}}, \bibinfo
  {author} {\bibfnamefont {D.~M.}\ \bibnamefont {Villeneuve}}, \ and\ \bibinfo
  {author} {\bibfnamefont {P.~B.}\ \bibnamefont {Corkum}},\ }\href {\doibase
  10.1088/0953-4075/42/18/185402} {\bibfield  {journal} {\bibinfo  {journal}
  {Journal of Physics B: Atomic, Molecular and Optical Physics}\ }\textbf
  {\bibinfo {volume} {42}},\ \bibinfo {pages} {185402} (\bibinfo {year}
  {2009})}\BibitemShut {NoStop}%
\bibitem [{\citenamefont {Hockett}\ \emph {et~al.}(2010)\citenamefont
  {Hockett}, \citenamefont {Staniforth},\ and\ \citenamefont
  {Reid}}]{hockett2010}%
  \BibitemOpen
  \bibfield  {author} {\bibinfo {author} {\bibfnamefont {P.}~\bibnamefont
  {Hockett}}, \bibinfo {author} {\bibfnamefont {M.}~\bibnamefont {Staniforth}},
  \ and\ \bibinfo {author} {\bibfnamefont {K.~L.}\ \bibnamefont {Reid}},\
  }\href {\doibase 10.1080/00268971003639266} {\bibfield  {journal} {\bibinfo
  {journal} {Molecular Physics}\ }\textbf {\bibinfo {volume} {108}},\ \bibinfo
  {pages} {1045} (\bibinfo {year} {2010})}\BibitemShut {NoStop}%
\bibitem [{\citenamefont {Wollenhaupt}\ \emph {et~al.}(2013)\citenamefont
  {Wollenhaupt}, \citenamefont {Lux}, \citenamefont {Krug},\ and\ \citenamefont
  {Baumert}}]{wollenhaupt2013}%
  \BibitemOpen
  \bibfield  {author} {\bibinfo {author} {\bibfnamefont {M.}~\bibnamefont
  {Wollenhaupt}}, \bibinfo {author} {\bibfnamefont {C.}~\bibnamefont {Lux}},
  \bibinfo {author} {\bibfnamefont {M.}~\bibnamefont {Krug}}, \ and\ \bibinfo
  {author} {\bibfnamefont {T.}~\bibnamefont {Baumert}},\ }\href {\doibase
  10.1002/cphc.201200968} {\bibfield  {journal} {\bibinfo  {journal}
  {ChemPhysChem}\ }\textbf {\bibinfo {volume} {14}},\ \bibinfo {pages} {1341}
  (\bibinfo {year} {2013})}\BibitemShut {NoStop}%
\bibitem [{\citenamefont {Askeland}\ \emph {et~al.}(2011)\citenamefont
  {Askeland}, \citenamefont {S\o{}rng\aa{}rd}, \citenamefont {Pilskog},
  \citenamefont {Nepstad},\ and\ \citenamefont {F\o{}rre}}]{askeland2011}%
  \BibitemOpen
  \bibfield  {author} {\bibinfo {author} {\bibfnamefont {S.}~\bibnamefont
  {Askeland}}, \bibinfo {author} {\bibfnamefont {S.~A.}\ \bibnamefont
  {S\o{}rng\aa{}rd}}, \bibinfo {author} {\bibfnamefont {I.}~\bibnamefont
  {Pilskog}}, \bibinfo {author} {\bibfnamefont {R.}~\bibnamefont {Nepstad}}, \
  and\ \bibinfo {author} {\bibfnamefont {M.}~\bibnamefont {F\o{}rre}},\ }\href
  {\doibase 10.1103/PhysRevA.84.033423} {\bibfield  {journal} {\bibinfo
  {journal} {Phys. Rev. A}\ }\textbf {\bibinfo {volume} {84}},\ \bibinfo
  {pages} {033423} (\bibinfo {year} {2011})}\BibitemShut {NoStop}%
\bibitem [{\citenamefont {Bauer}\ \emph {et~al.}(2014)\citenamefont {Bauer},
  \citenamefont {Mota-Furtado}, \citenamefont {O'Mahony}, \citenamefont
  {Piraux},\ and\ \citenamefont {Warda}}]{bauer2014}%
  \BibitemOpen
  \bibfield  {author} {\bibinfo {author} {\bibfnamefont {J.~H.}\ \bibnamefont
  {Bauer}}, \bibinfo {author} {\bibfnamefont {F.}~\bibnamefont {Mota-Furtado}},
  \bibinfo {author} {\bibfnamefont {P.~F.}\ \bibnamefont {O'Mahony}}, \bibinfo
  {author} {\bibfnamefont {B.}~\bibnamefont {Piraux}}, \ and\ \bibinfo {author}
  {\bibfnamefont {K.}~\bibnamefont {Warda}},\ }\href {\doibase
  10.1103/PhysRevA.90.063402} {\bibfield  {journal} {\bibinfo  {journal} {Phys.
  Rev. A}\ }\textbf {\bibinfo {volume} {90}},\ \bibinfo {pages} {063402}
  (\bibinfo {year} {2014})}\BibitemShut {NoStop}%
\bibitem [{\citenamefont {Wu}\ and\ \citenamefont {He}(2016)}]{wu2016}%
  \BibitemOpen
  \bibfield  {author} {\bibinfo {author} {\bibfnamefont {W.-Y.}\ \bibnamefont
  {Wu}}\ and\ \bibinfo {author} {\bibfnamefont {F.}~\bibnamefont {He}},\ }\href
  {\doibase 10.1103/PhysRevA.93.023415} {\bibfield  {journal} {\bibinfo
  {journal} {Phys. Rev. A}\ }\textbf {\bibinfo {volume} {93}},\ \bibinfo
  {pages} {023415} (\bibinfo {year} {2016})}\BibitemShut {NoStop}%
\bibitem [{\citenamefont {Ivanov}\ \emph {et~al.}(2017)\citenamefont {Ivanov},
  \citenamefont {Nam},\ and\ \citenamefont {Kim}}]{ivanov2017}%
  \BibitemOpen
  \bibfield  {author} {\bibinfo {author} {\bibfnamefont {I.~A.}\ \bibnamefont
  {Ivanov}}, \bibinfo {author} {\bibfnamefont {C.~H.}\ \bibnamefont {Nam}}, \
  and\ \bibinfo {author} {\bibfnamefont {K.~T.}\ \bibnamefont {Kim}},\ }\href
  {\doibase 10.1103/PhysRevA.95.053401} {\bibfield  {journal} {\bibinfo
  {journal} {Phys. Rev. A}\ }\textbf {\bibinfo {volume} {95}},\ \bibinfo
  {pages} {053401} (\bibinfo {year} {2017})}\BibitemShut {NoStop}%
\bibitem [{\citenamefont {Ngoko~Djiokap}\ \emph {et~al.}(2014)\citenamefont
  {Ngoko~Djiokap}, \citenamefont {Manakov}, \citenamefont {Meremianin},
  \citenamefont {Hu}, \citenamefont {Madsen},\ and\ \citenamefont
  {Starace}}]{djiokap2014}%
  \BibitemOpen
  \bibfield  {author} {\bibinfo {author} {\bibfnamefont {J.~M.}\ \bibnamefont
  {Ngoko~Djiokap}}, \bibinfo {author} {\bibfnamefont {N.~L.}\ \bibnamefont
  {Manakov}}, \bibinfo {author} {\bibfnamefont {A.~V.}\ \bibnamefont
  {Meremianin}}, \bibinfo {author} {\bibfnamefont {S.~X.}\ \bibnamefont {Hu}},
  \bibinfo {author} {\bibfnamefont {L.~B.}\ \bibnamefont {Madsen}}, \ and\
  \bibinfo {author} {\bibfnamefont {A.~F.}\ \bibnamefont {Starace}},\ }\href
  {\doibase 10.1103/PhysRevLett.113.223002} {\bibfield  {journal} {\bibinfo
  {journal} {Phys. Rev. Lett.}\ }\textbf {\bibinfo {volume} {113}},\ \bibinfo
  {pages} {223002} (\bibinfo {year} {2014})}\BibitemShut {NoStop}%
\bibitem [{\citenamefont {Ngoko~Djiokap}\ \emph {et~al.}(2016)\citenamefont
  {Ngoko~Djiokap}, \citenamefont {Meremianin}, \citenamefont {Manakov},
  \citenamefont {Hu}, \citenamefont {Madsen},\ and\ \citenamefont
  {Starace}}]{djiokap2016}%
  \BibitemOpen
  \bibfield  {author} {\bibinfo {author} {\bibfnamefont {J.~M.}\ \bibnamefont
  {Ngoko~Djiokap}}, \bibinfo {author} {\bibfnamefont {A.~V.}\ \bibnamefont
  {Meremianin}}, \bibinfo {author} {\bibfnamefont {N.~L.}\ \bibnamefont
  {Manakov}}, \bibinfo {author} {\bibfnamefont {S.~X.}\ \bibnamefont {Hu}},
  \bibinfo {author} {\bibfnamefont {L.~B.}\ \bibnamefont {Madsen}}, \ and\
  \bibinfo {author} {\bibfnamefont {A.~F.}\ \bibnamefont {Starace}},\ }\href
  {\doibase 10.1103/PhysRevA.94.013408} {\bibfield  {journal} {\bibinfo
  {journal} {Phys. Rev. A}\ }\textbf {\bibinfo {volume} {94}},\ \bibinfo
  {pages} {013408} (\bibinfo {year} {2016})}\BibitemShut {NoStop}%
\bibitem [{\citenamefont {Djiokap}\ \emph {et~al.}(2017)\citenamefont
  {Djiokap}, \citenamefont {Meremianin}, \citenamefont {Manakov}, \citenamefont
  {Hu}, \citenamefont {Madsen},\ and\ \citenamefont {Starace}}]{djiokap2017}%
  \BibitemOpen
  \bibfield  {author} {\bibinfo {author} {\bibfnamefont {J.~M.~N.}\
  \bibnamefont {Djiokap}}, \bibinfo {author} {\bibfnamefont {A.~V.}\
  \bibnamefont {Meremianin}}, \bibinfo {author} {\bibfnamefont {N.~L.}\
  \bibnamefont {Manakov}}, \bibinfo {author} {\bibfnamefont {S.~X.}\
  \bibnamefont {Hu}}, \bibinfo {author} {\bibfnamefont {L.~B.}\ \bibnamefont
  {Madsen}}, \ and\ \bibinfo {author} {\bibfnamefont {A.~F.}\ \bibnamefont
  {Starace}},\ }\href {\doibase 10.1103/PhysRevA.96.013405} {\bibfield
  {journal} {\bibinfo  {journal} {Phys. Rev. A}\ }\textbf {\bibinfo {volume}
  {96}},\ \bibinfo {pages} {013405} (\bibinfo {year} {2017})}\BibitemShut
  {NoStop}%
\bibitem [{\citenamefont {Donsa}\ \emph {et~al.}(2019)\citenamefont {Donsa},
  \citenamefont {B\ifmmode~\check{r}\else \v{r}\fi{}ezinov\'a}, \citenamefont
  {Ni}, \citenamefont {Feist},\ and\ \citenamefont {Burgd\"orfer}}]{donsa2019}%
  \BibitemOpen
  \bibfield  {author} {\bibinfo {author} {\bibfnamefont {S.}~\bibnamefont
  {Donsa}}, \bibinfo {author} {\bibfnamefont {I.}~\bibnamefont
  {B\ifmmode~\check{r}\else \v{r}\fi{}ezinov\'a}}, \bibinfo {author}
  {\bibfnamefont {H.}~\bibnamefont {Ni}}, \bibinfo {author} {\bibfnamefont
  {J.}~\bibnamefont {Feist}}, \ and\ \bibinfo {author} {\bibfnamefont
  {J.}~\bibnamefont {Burgd\"orfer}},\ }\href {\doibase
  10.1103/PhysRevA.99.023413} {\bibfield  {journal} {\bibinfo  {journal} {Phys.
  Rev. A}\ }\textbf {\bibinfo {volume} {99}},\ \bibinfo {pages} {023413}
  (\bibinfo {year} {2019})}\BibitemShut {NoStop}%
\bibitem [{\citenamefont {Ngoko~Djiokap}\ \emph {et~al.}(2018)\citenamefont
  {Ngoko~Djiokap}, \citenamefont {Meremianin}, \citenamefont {Manakov},
  \citenamefont {Madsen}, \citenamefont {Hu},\ and\ \citenamefont
  {Starace}}]{djiokap2018}%
  \BibitemOpen
  \bibfield  {author} {\bibinfo {author} {\bibfnamefont {J.~M.}\ \bibnamefont
  {Ngoko~Djiokap}}, \bibinfo {author} {\bibfnamefont {A.~V.}\ \bibnamefont
  {Meremianin}}, \bibinfo {author} {\bibfnamefont {N.~L.}\ \bibnamefont
  {Manakov}}, \bibinfo {author} {\bibfnamefont {L.~B.}\ \bibnamefont {Madsen}},
  \bibinfo {author} {\bibfnamefont {S.~X.}\ \bibnamefont {Hu}}, \ and\ \bibinfo
  {author} {\bibfnamefont {A.~F.}\ \bibnamefont {Starace}},\ }\href {\doibase
  10.1103/PhysRevA.98.063407} {\bibfield  {journal} {\bibinfo  {journal} {Phys.
  Rev. A}\ }\textbf {\bibinfo {volume} {98}},\ \bibinfo {pages} {063407}
  (\bibinfo {year} {2018})}\BibitemShut {NoStop}%
\bibitem [{\citenamefont {Pindzola}\ \emph {et~al.}(2017)\citenamefont
  {Pindzola}, \citenamefont {Li},\ and\ \citenamefont {Colgan}}]{pindzola2017}%
  \BibitemOpen
  \bibfield  {author} {\bibinfo {author} {\bibfnamefont {M.~S.}\ \bibnamefont
  {Pindzola}}, \bibinfo {author} {\bibfnamefont {Y.}~\bibnamefont {Li}}, \ and\
  \bibinfo {author} {\bibfnamefont {J.}~\bibnamefont {Colgan}},\ }\href
  {\doibase 10.1088/1361-6455/aa55f3} {\bibfield  {journal} {\bibinfo
  {journal} {Journal of Physics B: Atomic, Molecular and Optical Physics}\
  }\textbf {\bibinfo {volume} {50}},\ \bibinfo {pages} {045601} (\bibinfo
  {year} {2017})}\BibitemShut {NoStop}%
\bibitem [{\citenamefont {Yuan}\ \emph {et~al.}(2016)\citenamefont {Yuan},
  \citenamefont {Chelkowski},\ and\ \citenamefont {Bandrauk}}]{yuan2016}%
  \BibitemOpen
  \bibfield  {author} {\bibinfo {author} {\bibfnamefont {K.-J.}\ \bibnamefont
  {Yuan}}, \bibinfo {author} {\bibfnamefont {S.}~\bibnamefont {Chelkowski}}, \
  and\ \bibinfo {author} {\bibfnamefont {A.~D.}\ \bibnamefont {Bandrauk}},\
  }\href {\doibase 10.1103/PhysRevA.93.053425} {\bibfield  {journal} {\bibinfo
  {journal} {Phys. Rev. A}\ }\textbf {\bibinfo {volume} {93}},\ \bibinfo
  {pages} {053425} (\bibinfo {year} {2016})}\BibitemShut {NoStop}%
\bibitem [{\citenamefont {Yuan}\ \emph {et~al.}(2017)\citenamefont {Yuan},
  \citenamefont {Lu},\ and\ \citenamefont {Bandrauk}}]{yuan2017}%
  \BibitemOpen
  \bibfield  {author} {\bibinfo {author} {\bibfnamefont {K.-J.}\ \bibnamefont
  {Yuan}}, \bibinfo {author} {\bibfnamefont {H.}~\bibnamefont {Lu}}, \ and\
  \bibinfo {author} {\bibfnamefont {A.~D.}\ \bibnamefont {Bandrauk}},\ }\href
  {\doibase 10.1007/s11224-017-0964-5} {\bibfield  {journal} {\bibinfo
  {journal} {Structural Chemistry}\ }\textbf {\bibinfo {volume} {28}},\
  \bibinfo {pages} {1297} (\bibinfo {year} {2017})}\BibitemShut {NoStop}%
\bibitem [{\citenamefont {Yuan}\ and\ \citenamefont
  {Bandrauk}(2018)}]{yuan2018}%
  \BibitemOpen
  \bibfield  {author} {\bibinfo {author} {\bibfnamefont {K.-J.}\ \bibnamefont
  {Yuan}}\ and\ \bibinfo {author} {\bibfnamefont {A.~D.}\ \bibnamefont
  {Bandrauk}},\ }\href {\doibase 10.1103/PhysRevA.97.023408} {\bibfield
  {journal} {\bibinfo  {journal} {Phys. Rev. A}\ }\textbf {\bibinfo {volume}
  {97}},\ \bibinfo {pages} {023408} (\bibinfo {year} {2018})}\BibitemShut
  {NoStop}%
\bibitem [{\citenamefont {Averbukh}\ \emph {et~al.}(2001)\citenamefont
  {Averbukh}, \citenamefont {Alon},\ and\ \citenamefont
  {Moiseyev}}]{averbukh2001}%
  \BibitemOpen
  \bibfield  {author} {\bibinfo {author} {\bibfnamefont {V.}~\bibnamefont
  {Averbukh}}, \bibinfo {author} {\bibfnamefont {O.~E.}\ \bibnamefont {Alon}},
  \ and\ \bibinfo {author} {\bibfnamefont {N.}~\bibnamefont {Moiseyev}},\
  }\href {\doibase 10.1103/PhysRevA.64.033411} {\bibfield  {journal} {\bibinfo
  {journal} {Phys. Rev. A}\ }\textbf {\bibinfo {volume} {64}},\ \bibinfo
  {pages} {033411} (\bibinfo {year} {2001})}\BibitemShut {NoStop}%
\bibitem [{\citenamefont {Wardlow}\ and\ \citenamefont
  {Dundas}(2016)}]{wardlow2016}%
  \BibitemOpen
  \bibfield  {author} {\bibinfo {author} {\bibfnamefont {A.}~\bibnamefont
  {Wardlow}}\ and\ \bibinfo {author} {\bibfnamefont {D.}~\bibnamefont
  {Dundas}},\ }\href {\doibase 10.1103/PhysRevA.93.023428} {\bibfield
  {journal} {\bibinfo  {journal} {Phys. Rev. A}\ }\textbf {\bibinfo {volume}
  {93}},\ \bibinfo {pages} {023428} (\bibinfo {year} {2016})}\BibitemShut
  {NoStop}%
\bibitem [{\citenamefont {Kjeldsen}\ \emph {et~al.}(2007)\citenamefont
  {Kjeldsen}, \citenamefont {Nikolopoulos},\ and\ \citenamefont
  {Madsen}}]{kjeldsen2007}%
  \BibitemOpen
  \bibfield  {author} {\bibinfo {author} {\bibfnamefont {T.~K.}\ \bibnamefont
  {Kjeldsen}}, \bibinfo {author} {\bibfnamefont {L.~A.~A.}\ \bibnamefont
  {Nikolopoulos}}, \ and\ \bibinfo {author} {\bibfnamefont {L.~B.}\
  \bibnamefont {Madsen}},\ }\href {\doibase 10.1103/PhysRevA.75.063427}
  {\bibfield  {journal} {\bibinfo  {journal} {Phys. Rev. A}\ }\textbf {\bibinfo
  {volume} {75}},\ \bibinfo {pages} {063427} (\bibinfo {year}
  {2007})}\BibitemShut {NoStop}%
\bibitem [{\citenamefont {Abu-samha}\ and\ \citenamefont
  {Madsen}(2011)}]{abu2011}%
  \BibitemOpen
  \bibfield  {author} {\bibinfo {author} {\bibfnamefont {M.}~\bibnamefont
  {Abu-samha}}\ and\ \bibinfo {author} {\bibfnamefont {L.~B.}\ \bibnamefont
  {Madsen}},\ }\href {\doibase 10.1103/PhysRevA.84.023411} {\bibfield
  {journal} {\bibinfo  {journal} {Phys. Rev. A}\ }\textbf {\bibinfo {volume}
  {84}},\ \bibinfo {pages} {023411} (\bibinfo {year} {2011})}\BibitemShut
  {NoStop}%
\bibitem [{\citenamefont {Ivanov}\ and\ \citenamefont
  {Kheifets}(2013)}]{ivanov2013}%
  \BibitemOpen
  \bibfield  {author} {\bibinfo {author} {\bibfnamefont {I.~A.}\ \bibnamefont
  {Ivanov}}\ and\ \bibinfo {author} {\bibfnamefont {A.~S.}\ \bibnamefont
  {Kheifets}},\ }\href {\doibase 10.1103/PhysRevA.87.033407} {\bibfield
  {journal} {\bibinfo  {journal} {Phys. Rev. A}\ }\textbf {\bibinfo {volume}
  {87}},\ \bibinfo {pages} {033407} (\bibinfo {year} {2013})}\BibitemShut
  {NoStop}%
\bibitem [{\citenamefont {Ivanov}\ and\ \citenamefont
  {Kheifets}(2014)}]{ivanov2014}%
  \BibitemOpen
  \bibfield  {author} {\bibinfo {author} {\bibfnamefont {I.~A.}\ \bibnamefont
  {Ivanov}}\ and\ \bibinfo {author} {\bibfnamefont {A.~S.}\ \bibnamefont
  {Kheifets}},\ }\href {\doibase 10.1103/PhysRevA.89.021402} {\bibfield
  {journal} {\bibinfo  {journal} {Phys. Rev. A}\ }\textbf {\bibinfo {volume}
  {89}},\ \bibinfo {pages} {021402} (\bibinfo {year} {2014})}\BibitemShut
  {NoStop}%
\bibitem [{\citenamefont {Mancuso}\ \emph {et~al.}(2015)\citenamefont
  {Mancuso}, \citenamefont {Hickstein}, \citenamefont {Grychtol}, \citenamefont
  {Knut}, \citenamefont {Kfir}, \citenamefont {Tong}, \citenamefont {Dollar},
  \citenamefont {Zusin}, \citenamefont {Gopalakrishnan}, \citenamefont
  {Gentry}, \citenamefont {Turgut}, \citenamefont {Ellis}, \citenamefont
  {Chen}, \citenamefont {Fleischer}, \citenamefont {Cohen}, \citenamefont
  {Kapteyn},\ and\ \citenamefont {Murnane}}]{mancusco2015}%
  \BibitemOpen
  \bibfield  {author} {\bibinfo {author} {\bibfnamefont {C.~A.}\ \bibnamefont
  {Mancuso}}, \bibinfo {author} {\bibfnamefont {D.~D.}\ \bibnamefont
  {Hickstein}}, \bibinfo {author} {\bibfnamefont {P.}~\bibnamefont {Grychtol}},
  \bibinfo {author} {\bibfnamefont {R.}~\bibnamefont {Knut}}, \bibinfo {author}
  {\bibfnamefont {O.}~\bibnamefont {Kfir}}, \bibinfo {author} {\bibfnamefont
  {X.-M.}\ \bibnamefont {Tong}}, \bibinfo {author} {\bibfnamefont
  {F.}~\bibnamefont {Dollar}}, \bibinfo {author} {\bibfnamefont
  {D.}~\bibnamefont {Zusin}}, \bibinfo {author} {\bibfnamefont
  {M.}~\bibnamefont {Gopalakrishnan}}, \bibinfo {author} {\bibfnamefont
  {C.}~\bibnamefont {Gentry}}, \bibinfo {author} {\bibfnamefont
  {E.}~\bibnamefont {Turgut}}, \bibinfo {author} {\bibfnamefont {J.~L.}\
  \bibnamefont {Ellis}}, \bibinfo {author} {\bibfnamefont {M.-C.}\ \bibnamefont
  {Chen}}, \bibinfo {author} {\bibfnamefont {A.}~\bibnamefont {Fleischer}},
  \bibinfo {author} {\bibfnamefont {O.}~\bibnamefont {Cohen}}, \bibinfo
  {author} {\bibfnamefont {H.~C.}\ \bibnamefont {Kapteyn}}, \ and\ \bibinfo
  {author} {\bibfnamefont {M.~M.}\ \bibnamefont {Murnane}},\ }\href {\doibase
  10.1103/PhysRevA.91.031402} {\bibfield  {journal} {\bibinfo  {journal} {Phys.
  Rev. A}\ }\textbf {\bibinfo {volume} {91}},\ \bibinfo {pages} {031402}
  (\bibinfo {year} {2015})}\BibitemShut {NoStop}%
\bibitem [{\citenamefont {Ilchen}\ \emph {et~al.}(2017)\citenamefont {Ilchen},
  \citenamefont {Douguet}, \citenamefont {Mazza}, \citenamefont {Rafipoor},
  \citenamefont {Callegari}, \citenamefont {Finetti}, \citenamefont {Plekan},
  \citenamefont {Prince}, \citenamefont {Demidovich}, \citenamefont {Grazioli},
  \citenamefont {Avaldi}, \citenamefont {Bolognesi}, \citenamefont {Coreno},
  \citenamefont {Di~Fraia}, \citenamefont {Devetta}, \citenamefont
  {Ovcharenko}, \citenamefont {D\"usterer}, \citenamefont {Ueda}, \citenamefont
  {Bartschat}, \citenamefont {Grum-Grzhimailo}, \citenamefont {Bozhevolnov},
  \citenamefont {Kazansky}, \citenamefont {Kabachnik},\ and\ \citenamefont
  {Meyer}}]{ilchen2017}%
  \BibitemOpen
  \bibfield  {author} {\bibinfo {author} {\bibfnamefont {M.}~\bibnamefont
  {Ilchen}}, \bibinfo {author} {\bibfnamefont {N.}~\bibnamefont {Douguet}},
  \bibinfo {author} {\bibfnamefont {T.}~\bibnamefont {Mazza}}, \bibinfo
  {author} {\bibfnamefont {A.~J.}\ \bibnamefont {Rafipoor}}, \bibinfo {author}
  {\bibfnamefont {C.}~\bibnamefont {Callegari}}, \bibinfo {author}
  {\bibfnamefont {P.}~\bibnamefont {Finetti}}, \bibinfo {author} {\bibfnamefont
  {O.}~\bibnamefont {Plekan}}, \bibinfo {author} {\bibfnamefont {K.~C.}\
  \bibnamefont {Prince}}, \bibinfo {author} {\bibfnamefont {A.}~\bibnamefont
  {Demidovich}}, \bibinfo {author} {\bibfnamefont {C.}~\bibnamefont
  {Grazioli}}, \bibinfo {author} {\bibfnamefont {L.}~\bibnamefont {Avaldi}},
  \bibinfo {author} {\bibfnamefont {P.}~\bibnamefont {Bolognesi}}, \bibinfo
  {author} {\bibfnamefont {M.}~\bibnamefont {Coreno}}, \bibinfo {author}
  {\bibfnamefont {M.}~\bibnamefont {Di~Fraia}}, \bibinfo {author}
  {\bibfnamefont {M.}~\bibnamefont {Devetta}}, \bibinfo {author} {\bibfnamefont
  {Y.}~\bibnamefont {Ovcharenko}}, \bibinfo {author} {\bibfnamefont
  {S.}~\bibnamefont {D\"usterer}}, \bibinfo {author} {\bibfnamefont
  {K.}~\bibnamefont {Ueda}}, \bibinfo {author} {\bibfnamefont {K.}~\bibnamefont
  {Bartschat}}, \bibinfo {author} {\bibfnamefont {A.~N.}\ \bibnamefont
  {Grum-Grzhimailo}}, \bibinfo {author} {\bibfnamefont {A.~V.}\ \bibnamefont
  {Bozhevolnov}}, \bibinfo {author} {\bibfnamefont {A.~K.}\ \bibnamefont
  {Kazansky}}, \bibinfo {author} {\bibfnamefont {N.~M.}\ \bibnamefont
  {Kabachnik}}, \ and\ \bibinfo {author} {\bibfnamefont {M.}~\bibnamefont
  {Meyer}},\ }\href {\doibase 10.1103/PhysRevLett.118.013002} {\bibfield
  {journal} {\bibinfo  {journal} {Phys. Rev. Lett.}\ }\textbf {\bibinfo
  {volume} {118}},\ \bibinfo {pages} {013002} (\bibinfo {year}
  {2017})}\BibitemShut {NoStop}%
\bibitem [{\citenamefont {Weber}\ \emph {et~al.}(2000)\citenamefont {Weber},
  \citenamefont {Giessen}, \citenamefont {Weckenbrock}, \citenamefont
  {Urbasch}, \citenamefont {Staudte}, \citenamefont {Spielberger},
  \citenamefont {Jagutzki}, \citenamefont {Mergel}, \citenamefont {Vollmer},\
  and\ \citenamefont {D{\"o}rner}}]{weber2000}%
  \BibitemOpen
  \bibfield  {author} {\bibinfo {author} {\bibfnamefont {T.}~\bibnamefont
  {Weber}}, \bibinfo {author} {\bibfnamefont {H.}~\bibnamefont {Giessen}},
  \bibinfo {author} {\bibfnamefont {M.}~\bibnamefont {Weckenbrock}}, \bibinfo
  {author} {\bibfnamefont {G.}~\bibnamefont {Urbasch}}, \bibinfo {author}
  {\bibfnamefont {A.}~\bibnamefont {Staudte}}, \bibinfo {author} {\bibfnamefont
  {L.}~\bibnamefont {Spielberger}}, \bibinfo {author} {\bibfnamefont
  {O.}~\bibnamefont {Jagutzki}}, \bibinfo {author} {\bibfnamefont
  {V.}~\bibnamefont {Mergel}}, \bibinfo {author} {\bibfnamefont
  {M.}~\bibnamefont {Vollmer}}, \ and\ \bibinfo {author} {\bibfnamefont
  {R.}~\bibnamefont {D{\"o}rner}},\ }\href {\doibase 10.1038/35015033}
  {\bibfield  {journal} {\bibinfo  {journal} {Nature}\ }\textbf {\bibinfo
  {volume} {405}},\ \bibinfo {pages} {658} (\bibinfo {year}
  {2000})}\BibitemShut {NoStop}%
\bibitem [{\citenamefont {Drescher}\ \emph {et~al.}(2002)\citenamefont
  {Drescher}, \citenamefont {Hentschel}, \citenamefont {Kienberger},
  \citenamefont {Uiberacker}, \citenamefont {Yakovlev}, \citenamefont
  {Scrinzi}, \citenamefont {Westerwalbesloh}, \citenamefont {Kleineberg},
  \citenamefont {Heinzmann},\ and\ \citenamefont {Krausz}}]{drescher2002}%
  \BibitemOpen
  \bibfield  {author} {\bibinfo {author} {\bibfnamefont {M.}~\bibnamefont
  {Drescher}}, \bibinfo {author} {\bibfnamefont {M.}~\bibnamefont {Hentschel}},
  \bibinfo {author} {\bibfnamefont {R.}~\bibnamefont {Kienberger}}, \bibinfo
  {author} {\bibfnamefont {M.}~\bibnamefont {Uiberacker}}, \bibinfo {author}
  {\bibfnamefont {V.}~\bibnamefont {Yakovlev}}, \bibinfo {author}
  {\bibfnamefont {A.}~\bibnamefont {Scrinzi}}, \bibinfo {author} {\bibfnamefont
  {T.}~\bibnamefont {Westerwalbesloh}}, \bibinfo {author} {\bibfnamefont
  {U.}~\bibnamefont {Kleineberg}}, \bibinfo {author} {\bibfnamefont
  {U.}~\bibnamefont {Heinzmann}}, \ and\ \bibinfo {author} {\bibfnamefont
  {F.}~\bibnamefont {Krausz}},\ }\href {\doibase 10.1038/nature01143}
  {\bibfield  {journal} {\bibinfo  {journal} {Nature}\ }\textbf {\bibinfo
  {volume} {419}},\ \bibinfo {pages} {803} (\bibinfo {year}
  {2002})}\BibitemShut {NoStop}%
\bibitem [{\citenamefont {Uiberacker}\ \emph {et~al.}(2007)\citenamefont
  {Uiberacker}, \citenamefont {Uphues}, \citenamefont {Schultze}, \citenamefont
  {Verhoef}, \citenamefont {Yakovlev}, \citenamefont {Kling}, \citenamefont
  {Rauschenberger}, \citenamefont {Kabachnik}, \citenamefont {Schr{\"o}der},
  \citenamefont {Lezius}, \citenamefont {Kompa}, \citenamefont {Muller},
  \citenamefont {Vrakking}, \citenamefont {Hendel}, \citenamefont {Kleineberg},
  \citenamefont {Heinzmann}, \citenamefont {Drescher},\ and\ \citenamefont
  {Krausz}}]{uiberacker2007}%
  \BibitemOpen
  \bibfield  {author} {\bibinfo {author} {\bibfnamefont {M.}~\bibnamefont
  {Uiberacker}}, \bibinfo {author} {\bibfnamefont {T.}~\bibnamefont {Uphues}},
  \bibinfo {author} {\bibfnamefont {M.}~\bibnamefont {Schultze}}, \bibinfo
  {author} {\bibfnamefont {A.~J.}\ \bibnamefont {Verhoef}}, \bibinfo {author}
  {\bibfnamefont {V.}~\bibnamefont {Yakovlev}}, \bibinfo {author}
  {\bibfnamefont {M.~F.}\ \bibnamefont {Kling}}, \bibinfo {author}
  {\bibfnamefont {J.}~\bibnamefont {Rauschenberger}}, \bibinfo {author}
  {\bibfnamefont {N.~M.}\ \bibnamefont {Kabachnik}}, \bibinfo {author}
  {\bibfnamefont {H.}~\bibnamefont {Schr{\"o}der}}, \bibinfo {author}
  {\bibfnamefont {M.}~\bibnamefont {Lezius}}, \bibinfo {author} {\bibfnamefont
  {K.~L.}\ \bibnamefont {Kompa}}, \bibinfo {author} {\bibfnamefont {H.~G.}\
  \bibnamefont {Muller}}, \bibinfo {author} {\bibfnamefont {M.~J.~J.}\
  \bibnamefont {Vrakking}}, \bibinfo {author} {\bibfnamefont {S.}~\bibnamefont
  {Hendel}}, \bibinfo {author} {\bibfnamefont {U.}~\bibnamefont {Kleineberg}},
  \bibinfo {author} {\bibfnamefont {U.}~\bibnamefont {Heinzmann}}, \bibinfo
  {author} {\bibfnamefont {M.}~\bibnamefont {Drescher}}, \ and\ \bibinfo
  {author} {\bibfnamefont {F.}~\bibnamefont {Krausz}},\ }\href
  {https://doi.org/10.1038/nature05648} {\bibfield  {journal} {\bibinfo
  {journal} {Nature}\ }\textbf {\bibinfo {volume} {446}},\ \bibinfo {pages}
  {627} (\bibinfo {year} {2007})}\BibitemShut {NoStop}%
\bibitem [{\citenamefont {Leone}\ \emph {et~al.}(2014)\citenamefont {Leone},
  \citenamefont {McCurdy}, \citenamefont {Burgd{\"o}rfer}, \citenamefont
  {Cederbaum}, \citenamefont {Chang}, \citenamefont {Dudovich}, \citenamefont
  {Feist}, \citenamefont {Greene}, \citenamefont {Ivanov}, \citenamefont
  {Kienberger}, \citenamefont {Keller}, \citenamefont {Kling}, \citenamefont
  {Loh}, \citenamefont {Pfeifer}, \citenamefont {Pfeiffer}, \citenamefont
  {Santra}, \citenamefont {Schafer}, \citenamefont {Stolow}, \citenamefont
  {Thumm},\ and\ \citenamefont {Vrakking}}]{leone2014}%
  \BibitemOpen
  \bibfield  {author} {\bibinfo {author} {\bibfnamefont {S.~R.}\ \bibnamefont
  {Leone}}, \bibinfo {author} {\bibfnamefont {C.~W.}\ \bibnamefont {McCurdy}},
  \bibinfo {author} {\bibfnamefont {J.}~\bibnamefont {Burgd{\"o}rfer}},
  \bibinfo {author} {\bibfnamefont {L.~S.}\ \bibnamefont {Cederbaum}}, \bibinfo
  {author} {\bibfnamefont {Z.}~\bibnamefont {Chang}}, \bibinfo {author}
  {\bibfnamefont {N.}~\bibnamefont {Dudovich}}, \bibinfo {author}
  {\bibfnamefont {J.}~\bibnamefont {Feist}}, \bibinfo {author} {\bibfnamefont
  {C.~H.}\ \bibnamefont {Greene}}, \bibinfo {author} {\bibfnamefont
  {M.}~\bibnamefont {Ivanov}}, \bibinfo {author} {\bibfnamefont
  {R.}~\bibnamefont {Kienberger}}, \bibinfo {author} {\bibfnamefont
  {U.}~\bibnamefont {Keller}}, \bibinfo {author} {\bibfnamefont {M.~F.}\
  \bibnamefont {Kling}}, \bibinfo {author} {\bibfnamefont {Z.-H.}\ \bibnamefont
  {Loh}}, \bibinfo {author} {\bibfnamefont {T.}~\bibnamefont {Pfeifer}},
  \bibinfo {author} {\bibfnamefont {A.~N.}\ \bibnamefont {Pfeiffer}}, \bibinfo
  {author} {\bibfnamefont {R.}~\bibnamefont {Santra}}, \bibinfo {author}
  {\bibfnamefont {K.}~\bibnamefont {Schafer}}, \bibinfo {author} {\bibfnamefont
  {A.}~\bibnamefont {Stolow}}, \bibinfo {author} {\bibfnamefont
  {U.}~\bibnamefont {Thumm}}, \ and\ \bibinfo {author} {\bibfnamefont
  {M.~J.~J.}\ \bibnamefont {Vrakking}},\ }\href
  {https://doi.org/10.1038/nphoton.2014.48} {\bibfield  {journal} {\bibinfo
  {journal} {Nature Photonics}\ }\textbf {\bibinfo {volume} {8}},\ \bibinfo
  {pages} {162} (\bibinfo {year} {2014})}\BibitemShut {NoStop}%
\bibitem [{\citenamefont {Nikolopoulos}\ \emph {et~al.}(2008)\citenamefont
  {Nikolopoulos}, \citenamefont {Parker},\ and\ \citenamefont
  {Taylor}}]{lampros2008}%
  \BibitemOpen
  \bibfield  {author} {\bibinfo {author} {\bibfnamefont {L.~A.~A.}\
  \bibnamefont {Nikolopoulos}}, \bibinfo {author} {\bibfnamefont {J.~S.}\
  \bibnamefont {Parker}}, \ and\ \bibinfo {author} {\bibfnamefont {K.~T.}\
  \bibnamefont {Taylor}},\ }\href {\doibase 10.1103/PhysRevA.78.063420}
  {\bibfield  {journal} {\bibinfo  {journal} {Phys. Rev. A}\ }\textbf {\bibinfo
  {volume} {78}},\ \bibinfo {pages} {063420} (\bibinfo {year}
  {2008})}\BibitemShut {NoStop}%
\bibitem [{\citenamefont {Moore}\ \emph {et~al.}(2011)\citenamefont {Moore},
  \citenamefont {Lysaght}, \citenamefont {Nikolopoulos}, \citenamefont
  {Parker}, \citenamefont {van~der Hart},\ and\ \citenamefont
  {Taylor}}]{moore2011}%
  \BibitemOpen
  \bibfield  {author} {\bibinfo {author} {\bibfnamefont {L.}~\bibnamefont
  {Moore}}, \bibinfo {author} {\bibfnamefont {M.}~\bibnamefont {Lysaght}},
  \bibinfo {author} {\bibfnamefont {L.}~\bibnamefont {Nikolopoulos}}, \bibinfo
  {author} {\bibfnamefont {J.}~\bibnamefont {Parker}}, \bibinfo {author}
  {\bibfnamefont {H.}~\bibnamefont {van~der Hart}}, \ and\ \bibinfo {author}
  {\bibfnamefont {K.}~\bibnamefont {Taylor}},\ }\href {\doibase
  10.1080/09500340.2011.559315} {\bibfield  {journal} {\bibinfo  {journal}
  {Journal of Modern Optics}\ }\textbf {\bibinfo {volume} {58}},\ \bibinfo
  {pages} {1132} (\bibinfo {year} {2011})}\BibitemShut {NoStop}%
\bibitem [{\citenamefont {Clarke}\ \emph
  {et~al.}(2018{\natexlab{a}})\citenamefont {Clarke}, \citenamefont
  {Armstrong}, \citenamefont {Brown},\ and\ \citenamefont {van~der
  Hart}}]{clarke2018}%
  \BibitemOpen
  \bibfield  {author} {\bibinfo {author} {\bibfnamefont {D.~D.~A.}\
  \bibnamefont {Clarke}}, \bibinfo {author} {\bibfnamefont {G.~S.~J.}\
  \bibnamefont {Armstrong}}, \bibinfo {author} {\bibfnamefont {A.~C.}\
  \bibnamefont {Brown}}, \ and\ \bibinfo {author} {\bibfnamefont {H.~W.}\
  \bibnamefont {van~der Hart}},\ }\href {\doibase 10.1103/PhysRevA.98.053442}
  {\bibfield  {journal} {\bibinfo  {journal} {Phys. Rev. A}\ }\textbf {\bibinfo
  {volume} {98}},\ \bibinfo {pages} {053442} (\bibinfo {year}
  {2018}{\natexlab{a}})}\BibitemShut {NoStop}%
\bibitem [{\citenamefont {Brown}\ \emph {et~al.}(2019)\citenamefont {Brown},
  \citenamefont {Armstrong}, \citenamefont {Benda}, \citenamefont {Clarke},
  \citenamefont {Wragg}, \citenamefont {Hamilton}, \citenamefont
  {Ma\v{s}\'{i}n}, \citenamefont {Gorfinkiel},\ and\ \citenamefont {van~der
  Hart}}]{brown2019}%
  \BibitemOpen
  \bibfield  {author} {\bibinfo {author} {\bibfnamefont {A.~C.}\ \bibnamefont
  {Brown}}, \bibinfo {author} {\bibfnamefont {G.~S.~J.}\ \bibnamefont
  {Armstrong}}, \bibinfo {author} {\bibfnamefont {J.}~\bibnamefont {Benda}},
  \bibinfo {author} {\bibfnamefont {D.~D.~A.}\ \bibnamefont {Clarke}}, \bibinfo
  {author} {\bibfnamefont {J.}~\bibnamefont {Wragg}}, \bibinfo {author}
  {\bibfnamefont {K.~R.}\ \bibnamefont {Hamilton}}, \bibinfo {author}
  {\bibfnamefont {Z.}~\bibnamefont {Ma\v{s}\'{i}n}}, \bibinfo {author}
  {\bibfnamefont {J.}~\bibnamefont {Gorfinkiel}}, \ and\ \bibinfo {author}
  {\bibfnamefont {H.~W.}\ \bibnamefont {van~der Hart}},\ }\href@noop {}
  {\bibfield  {journal} {\bibinfo  {journal} {Computer Physics Communications}\
  }\textbf {\bibinfo {volume} {XX}},\ \bibinfo {pages} {YYYY} (\bibinfo {year}
  {2019})}\BibitemShut {NoStop}%
\bibitem [{\citenamefont {Hassouneh}\ \emph {et~al.}(2014)\citenamefont
  {Hassouneh}, \citenamefont {Brown},\ and\ \citenamefont {van~der
  Hart}}]{hassouneh2014}%
  \BibitemOpen
  \bibfield  {author} {\bibinfo {author} {\bibfnamefont {O.}~\bibnamefont
  {Hassouneh}}, \bibinfo {author} {\bibfnamefont {A.~C.}\ \bibnamefont
  {Brown}}, \ and\ \bibinfo {author} {\bibfnamefont {H.~W.}\ \bibnamefont
  {van~der Hart}},\ }\href {\doibase 10.1103/PhysRevA.90.043418} {\bibfield
  {journal} {\bibinfo  {journal} {Phys. Rev. A}\ }\textbf {\bibinfo {volume}
  {90}},\ \bibinfo {pages} {043418} (\bibinfo {year} {2014})}\BibitemShut
  {NoStop}%
\bibitem [{\citenamefont {Hassouneh}\ \emph {et~al.}(2015)\citenamefont
  {Hassouneh}, \citenamefont {Law}, \citenamefont {Shearer}, \citenamefont
  {Brown},\ and\ \citenamefont {van~der Hart}}]{hassouneh2015}%
  \BibitemOpen
  \bibfield  {author} {\bibinfo {author} {\bibfnamefont {O.}~\bibnamefont
  {Hassouneh}}, \bibinfo {author} {\bibfnamefont {S.}~\bibnamefont {Law}},
  \bibinfo {author} {\bibfnamefont {S.~F.~C.}\ \bibnamefont {Shearer}},
  \bibinfo {author} {\bibfnamefont {A.~C.}\ \bibnamefont {Brown}}, \ and\
  \bibinfo {author} {\bibfnamefont {H.~W.}\ \bibnamefont {van~der Hart}},\
  }\href {\doibase 10.1103/PhysRevA.91.031404} {\bibfield  {journal} {\bibinfo
  {journal} {Phys. Rev. A}\ }\textbf {\bibinfo {volume} {91}},\ \bibinfo
  {pages} {031404} (\bibinfo {year} {2015})}\BibitemShut {NoStop}%
\bibitem [{\citenamefont {Brown}\ and\ \citenamefont {van~der
  Hart}(2016)}]{brown2016}%
  \BibitemOpen
  \bibfield  {author} {\bibinfo {author} {\bibfnamefont {A.~C.}\ \bibnamefont
  {Brown}}\ and\ \bibinfo {author} {\bibfnamefont {H.~W.}\ \bibnamefont
  {van~der Hart}},\ }\href {\doibase 10.1103/PhysRevLett.117.093201} {\bibfield
   {journal} {\bibinfo  {journal} {Phys. Rev. Lett.}\ }\textbf {\bibinfo
  {volume} {117}},\ \bibinfo {pages} {093201} (\bibinfo {year}
  {2016})}\BibitemShut {NoStop}%
\bibitem [{\citenamefont {Clarke}\ \emph
  {et~al.}(2018{\natexlab{b}})\citenamefont {Clarke}, \citenamefont {van~der
  Hart},\ and\ \citenamefont {Brown}}]{clarke2017}%
  \BibitemOpen
  \bibfield  {author} {\bibinfo {author} {\bibfnamefont {D.~D.~A.}\
  \bibnamefont {Clarke}}, \bibinfo {author} {\bibfnamefont {H.~W.}\
  \bibnamefont {van~der Hart}}, \ and\ \bibinfo {author} {\bibfnamefont
  {A.~C.}\ \bibnamefont {Brown}},\ }\href {\doibase 10.1103/PhysRevA.97.023413}
  {\bibfield  {journal} {\bibinfo  {journal} {Phys. Rev. A}\ }\textbf {\bibinfo
  {volume} {97}},\ \bibinfo {pages} {023413} (\bibinfo {year}
  {2018}{\natexlab{b}})}\BibitemShut {NoStop}%
\bibitem [{\citenamefont {Armstrong}\ \emph {et~al.}(2019)\citenamefont
  {Armstrong}, \citenamefont {Clarke}, \citenamefont {Brown},\ and\
  \citenamefont {van~der Hart}}]{armstrong2019}%
  \BibitemOpen
  \bibfield  {author} {\bibinfo {author} {\bibfnamefont {G.~S.~J.}\
  \bibnamefont {Armstrong}}, \bibinfo {author} {\bibfnamefont {D.~D.~A.}\
  \bibnamefont {Clarke}}, \bibinfo {author} {\bibfnamefont {A.~C.}\
  \bibnamefont {Brown}}, \ and\ \bibinfo {author} {\bibfnamefont {H.~W.}\
  \bibnamefont {van~der Hart}},\ }\href {\doibase 10.1103/PhysRevA.99.023429}
  {\bibfield  {journal} {\bibinfo  {journal} {Phys. Rev. A}\ }\textbf {\bibinfo
  {volume} {99}},\ \bibinfo {pages} {023429} (\bibinfo {year}
  {2019})}\BibitemShut {NoStop}%
\bibitem [{\citenamefont {Hutchinson}\ \emph {et~al.}(2010)\citenamefont
  {Hutchinson}, \citenamefont {Lysaght},\ and\ \citenamefont {van~der
  Hart}}]{hutchinson2010}%
  \BibitemOpen
  \bibfield  {author} {\bibinfo {author} {\bibfnamefont {S.}~\bibnamefont
  {Hutchinson}}, \bibinfo {author} {\bibfnamefont {M.~A.}\ \bibnamefont
  {Lysaght}}, \ and\ \bibinfo {author} {\bibfnamefont {H.~W.}\ \bibnamefont
  {van~der Hart}},\ }\href {\doibase 10.1088/0953-4075/43/9/095603} {\bibfield
  {journal} {\bibinfo  {journal} {Journal of Physics B: Atomic, Molecular and
  Optical Physics}\ }\textbf {\bibinfo {volume} {43}},\ \bibinfo {pages}
  {095603} (\bibinfo {year} {2010})}\BibitemShut {NoStop}%
\bibitem [{\citenamefont {Clementi}\ and\ \citenamefont
  {Roetti}(1974)}]{clemroe}%
  \BibitemOpen
  \bibfield  {author} {\bibinfo {author} {\bibfnamefont {E.}~\bibnamefont
  {Clementi}}\ and\ \bibinfo {author} {\bibfnamefont {C.}~\bibnamefont
  {Roetti}},\ }\href {\doibase https://doi.org/10.1016/S0092-640X(74)80016-1}
  {\bibfield  {journal} {\bibinfo  {journal} {Atomic Data and Nuclear Data
  Tables}\ }\textbf {\bibinfo {volume} {14}},\ \bibinfo {pages} {177 }
  (\bibinfo {year} {1974})}\BibitemShut {NoStop}%
\bibitem [{\citenamefont {Berrington}\ \emph {et~al.}(2006)\citenamefont
  {Berrington}, \citenamefont {Nakazaki},\ and\ \citenamefont
  {Murakami}}]{berrington2006}%
  \BibitemOpen
  \bibfield  {author} {\bibinfo {author} {\bibfnamefont {K.~A.}\ \bibnamefont
  {Berrington}}, \bibinfo {author} {\bibfnamefont {S.}~\bibnamefont
  {Nakazaki}}, \ and\ \bibinfo {author} {\bibfnamefont {Y.}~\bibnamefont
  {Murakami}},\ }\href {\doibase 10.1088/0953-4075/39/22/016} {\bibfield
  {journal} {\bibinfo  {journal} {Journal of Physics B: Atomic, Molecular and
  Optical Physics}\ }\textbf {\bibinfo {volume} {39}},\ \bibinfo {pages} {4733}
  (\bibinfo {year} {2006})}\BibitemShut {NoStop}%
\bibitem [{\citenamefont {Hibbert}(1975)}]{hibbert1975}%
  \BibitemOpen
  \bibfield  {author} {\bibinfo {author} {\bibfnamefont {A.}~\bibnamefont
  {Hibbert}},\ }\href {\doibase https://doi.org/10.1016/0010-4655(75)90103-4}
  {\bibfield  {journal} {\bibinfo  {journal} {Computer Physics Communications}\
  }\textbf {\bibinfo {volume} {9}},\ \bibinfo {pages} {141 } (\bibinfo {year}
  {1975})}\BibitemShut {NoStop}%
\bibitem [{\citenamefont {Kramida}\ \emph {et~al.}(2018)\citenamefont
  {Kramida}, \citenamefont {{Yu.~Ralchenko}}, \citenamefont {Reader},\ and\
  \citenamefont {{NIST ASD Team}}}]{nist}%
  \BibitemOpen
  \bibfield  {author} {\bibinfo {author} {\bibfnamefont {A.}~\bibnamefont
  {Kramida}}, \bibinfo {author} {\bibnamefont {{Yu.~Ralchenko}}}, \bibinfo
  {author} {\bibfnamefont {J.}~\bibnamefont {Reader}}, \ and\ \bibinfo {author}
  {\bibnamefont {{NIST ASD Team}}},\ }\href@noop {} {}\bibinfo {howpublished}
  {{NIST Atomic Spectra Database (ver. 5.6.1). Available:
  {\url{https://physics.nist.gov/asd}}. National Institute of Standards and
  Technology, Gaithersburg, MD.}} (\bibinfo {year} {2018})\BibitemShut
  {NoStop}%
\bibitem [{\citenamefont {van~der Hart}\ \emph {et~al.}(2008)\citenamefont
  {van~der Hart}, \citenamefont {Lysaght},\ and\ \citenamefont
  {Burke}}]{vdh2008}%
  \BibitemOpen
  \bibfield  {author} {\bibinfo {author} {\bibfnamefont {H.~W.}\ \bibnamefont
  {van~der Hart}}, \bibinfo {author} {\bibfnamefont {M.~A.}\ \bibnamefont
  {Lysaght}}, \ and\ \bibinfo {author} {\bibfnamefont {P.~G.}\ \bibnamefont
  {Burke}},\ }\href {\doibase 10.1103/PhysRevA.77.065401} {\bibfield  {journal}
  {\bibinfo  {journal} {Phys. Rev. A}\ }\textbf {\bibinfo {volume} {77}},\
  \bibinfo {pages} {065401} (\bibinfo {year} {2008})}\BibitemShut {NoStop}%
\bibitem [{\citenamefont {Lysaght}\ \emph {et~al.}(2009)\citenamefont
  {Lysaght}, \citenamefont {van~der Hart},\ and\ \citenamefont
  {Burke}}]{lysaght2009}%
  \BibitemOpen
  \bibfield  {author} {\bibinfo {author} {\bibfnamefont {M.~A.}\ \bibnamefont
  {Lysaght}}, \bibinfo {author} {\bibfnamefont {H.~W.}\ \bibnamefont {van~der
  Hart}}, \ and\ \bibinfo {author} {\bibfnamefont {P.~G.}\ \bibnamefont
  {Burke}},\ }\href {\doibase 10.1103/PhysRevA.79.053411} {\bibfield  {journal}
  {\bibinfo  {journal} {Phys. Rev. A}\ }\textbf {\bibinfo {volume} {79}},\
  \bibinfo {pages} {053411} (\bibinfo {year} {2009})}\BibitemShut {NoStop}%
\bibitem [{pur(2019)}]{pure}%
  \BibitemOpen
  \href@noop {} {}\bibinfo {howpublished} {PURE research database
  {\url{https://doi.org/10.17034/24d396fc-1744-4431-939d-0658252a854b}}}
  (\bibinfo {year} {2019})\BibitemShut {NoStop}%
\bibitem [{rep(2019)}]{repo}%
  \BibitemOpen
  \href@noop {} {}\bibinfo {howpublished} {The RMT repository
  {\url{https://gitlab.com/Uk-amor/RMT/rmt}}} (\bibinfo {year}
  {2019})\BibitemShut {NoStop}%
\bibitem [{\citenamefont {Fano}\ and\ \citenamefont {Racah}(1959)}]{fano1959}%
  \BibitemOpen
  \bibfield  {author} {\bibinfo {author} {\bibfnamefont {U.}~\bibnamefont
  {Fano}}\ and\ \bibinfo {author} {\bibfnamefont {G.}~\bibnamefont {Racah}},\
  }\href {https://books.google.co.uk/books?id=buBEAAAAIAAJ} {\emph {\bibinfo
  {title} {Irreducible tensorial sets}}},\ Pure and applied physics\ (\bibinfo
  {publisher} {Academic Press},\ \bibinfo {year} {1959})\BibitemShut {NoStop}%
\end{thebibliography}%

\end{document}